\documentclass[10pt]{iopart}

\usepackage{iopams}
\usepackage{cite}
\usepackage{graphicx}  
\usepackage{float}
\usepackage{braket}
\usepackage[table,xcdraw]{xcolor}
\usepackage{comment}
\usepackage[caption=false]{subfig}
\graphicspath{{images/}}
\begin{document}

\title[Dimensional analysis and quantum-classical correspondence]{Dimensional analysis and the correspondence between
classical and quantum uncertainty}

\author{V Gattus$^1$ and S Karamitsos$^2$}
\address{School of Physics and Astronomy, University of Manchester, \\
Manchester M13 9PL, United Kingdom}
\eads{\mailto{$^1$ viola.gattus@student.manchester.ac.uk}, \mailto{ $^2 $sotirios.karamitsos@manchester.ac.uk}}
\vspace{10pt}
\begin{indented}
\item[]May 2020
\end{indented}

\begin{abstract}
Heisenberg's uncertainty principle is often cited as an example of a ``purely quantum'' relation with no analogue in the classical limit where $\hbar \to 0$. However, this formulation of the classical limit is problematic for many reasons, one of which is dimensional analysis. Since $\hbar$ is a dimensionful constant, we may always work in natural units in which $\hbar = 1$. Dimensional analysis teaches us that all physical laws can be expressed purely in terms of dimensionless quantities. This  indicates that the existence of a dimensionally consistent constraint on $\Delta x \Delta p$ requires the existence of a dimensionful parameter with  units of action, and that any definition of the classical limit must be formulated in terms of dimensionless quantities (such as quantum numbers). Therefore, bounds on classical uncertainty (formulated in terms of statistical ensembles) can only be written in terms of dimensionful scales of the system under consideration, and can be readily compared to their quantum counterparts after being non-dimensionalized. We compare the uncertainty of certain coupled classical systems and their quantum counterparts (such as harmonic oscillators and particles in a box), and show that they converge in the classical limit. We find that since these systems feature additional dimensionful scales, the uncertainty bounds are dependent on multiple dimensionless parameters, in accordance with dimensional considerations.
\end{abstract}

%
\vspace{2pc}
\noindent{\it Keywords}: classical limit, dimensional analysis, quantum uncertainty
%
%
%

\section{Introduction}

Historically, the development of quantum mechanics benefited greatly from the intuition offered by the Lagrangian and Hamiltonian formulations of classical mechanics. Perhaps the most famous example of this was Schr\"odinger's derivation of his eponymous equation. Realizing that microscopic particles had a fundamentally ``undulatory'' (wave-like) nature, Schr\"odinger observed that constant action surfaces\footnote{Here,  ``action" refers to $W = \int L dt$, where $L$ is the Lagrangian.}, being periodic, would play a major role in developing a robust theory of quantum mechanics \cite{schrodinger}. It is at this point in which dimensional analysis played a crucial role in Schr\"odinger's reasoning: indeed, since the action is a dimensionful quantity, it can only appear as the argument of a periodic function if it is divided by a quantity that also has units of action. If one cannot ``resist the temptation'', as Schr\"odinger put it, of assuming that this constant is a fundamental constant, the most obvious candidate is the Planck constant $h$, which can be converted into a phase if divided by $2\pi$, leading to the appearance of the reduced Planck constant $\hbar \equiv h/2\pi$.

Schr\"odinger's archetypal quantization procedure was later formalized and refined as the deformation from Poisson brackets to quantum commutators and the promotion of classical expressions to quantum operators for classical systems. It is therefore natural to ask whether ``purely quantum" phenomena can be formulated in a classical manner. The question has been approached from multiple angles, and, while it may be futile to attempt to build a purely classical theory that perfectly replicates the observations of quantum mechanics, we at least expect quantum mechanics to reproduce classical predictions corresponding to our everyday experience. This is the crux of the \emph{correspondence principle} \cite{bernal,huang,liboff,cabrera}.

The correspondence principle is often expressed as the idea that quantum mechanics must reduce to classical mechanics in some limit, usually referred to  as the \emph{classical limit}. Stated like this, the correspondence principle is little more than tautological unless we are very careful to expand on how the classical limit is formally defined. In most undergraduate textbooks, the correspondence principle is formalized in terms of Ehrenfest's theorem, which relates the evolution of the expectation values of operators to their commutator with the Hamiltonian. However, even Ehrenfest's theorem alone does not bridge the conceptual gap between quantum and classical mechanics \cite{ballentine}. This is only possible if we assign a classical meaning to the expectation values themselves. 
In order to do this, consider a quantum system evolving so rapidly that we can only measure its mean state over many cycles. Hence, the classical limit may be identified as the limit in which energies (or quantum numbers) are so large that the law of large numbers is unquestionably in effect and our macroscopic measurements are, in effect, averages of a very large number of quantum measurements.

While Ehrenfest's theorem provides a formal way of moving from the quantum realm of uncertainty to the classical realm of certainty, it is helpful to approach the question from the opposite angle and ask how uncertainty may arise in classical systems. Of course, for a single system with a fixed, known state, the notion of an expectation value of a quantity is no different from the quantity itself. However, since quantum mechanics is an inherently probabilistic theory, it makes more sense to formulate classical uncertainty by way of an \emph{ensemble} of systems, each with a different statistical weight. In this framework, the classical probability distribution is the analogue of the quantum mechanical probability density $\rho = |\psi|^2$. Therefore, in this way, the statistical average of the ensemble may be identified as the classical analogue of the quantum average of the system in question. 

We turn our attention to the well-known Heisenberg uncertainty principle, a purely quantum mechanical limit on the simultaneous localization of a particle's energy and momentum $\Delta x \:  \Delta p \:  \ge  \:   \hbar/2$. A na\"ive treatment of the classical limit might lead us to expect that no uncertainty exists in the classical case. However, this would be a mistake: the Heisenberg uncertainty principle presupposes the existence of a fundamental unit of action (i.e. the Planck constant $h$), something which does not exist in the classical realm. Therefore, a meaningful comparison between the two domains can only be achieved by working with non-dimensionalized quantities, such as the quantum numbers, recognized by Schr\"odinger as corresponding to Bohr's stationary energy levels of the elliptic orbits in the case of the atom. Schr\"odinger even obliquely refers to the correspondence principle at this stage, noting that even though ``mathematically", the atom spreads throughout space as $n\to\infty$, its quantum nature is restricted to a few angstroms, i.e. for relatively small $n$ \cite{schrodinger}. 

The idea that a comparison of the quantum and classical realm must be made through the use of pure numbers is by no means exclusive to the correspondence principle: it is a result of the Buckingham-$\pi$ theorem \cite{buckingham}, which states that \emph{every} law of physics must be expressible in a non-dimensionalized form. A careful treatment of the harmonic oscillator, infinite square well, and other single-particle systems with simple one-dimensional potentials \cite{doncheski,korsch,singh,kheiri,robinett5,radozycki,doncheski2,yoder} reveals that, once position and momentum are properly non-dimensionalized (with the help of $x_0$ and $p_0$, which are reference scales inherent to the system), then it is possible to arrive at a similar limit for the ensemble uncertainty, given by $\Delta \bar x  \:  \Delta \bar p \:  \ge c $, where $c$ is a pure number.

A natural question to ask at this stage is how the relation between quantum and classical uncertainty is modified for interacting one dimensional systems with additional dimensionful scales (such as coupled systems). Our approach is to define appropriate dimensionless variables, and seek to obtain the uncertainty relations in terms of their dimensionless ratios. In particular, we expect that adding more degrees of freedom to the system will yield more dimensionless ratios in the expression for the uncertainty relations, in accordance with the Buckingham-$\pi$ theorem.

The outline of the paper is as follows: in Section 2, we briefly review dimensional analysis
and consider the implications of the Buckingham-$\pi$ theorem for the classical and quantum uncertainty principle. In Section 3, we outline how the notion of classical uncertainty may be formalized by providing an overview of the correspondence between classical and quantum uncertainty for one-dimensional harmonic oscillators and the infinite square well. We then go on to examine how the addition of further dimensionful scales modifies the uncertainty relations while still respecting the correspondence principle. In Sections 4 and 5, we study coupled harmonic oscillators, whereas in Section 6, we study two particles in a box with a contact potential. Finally, we discuss our findings in the Conclusions.

\section{Dimensional analysis and the classical limit}
Physical dimensions are a concept related to but distinct from that of units.  Units are related to \emph{standards} (e.g. the length of a rod, the weight of a bearing, etc.) and correspond to the way in which the same measurements can be expressed:  for instance, $1 \ \rm{km}$ and $1000 \ \rm{m}$ both describe the same quantity. Dimensions, however, are inherent qualities of a physical quantity, related to how they are measured within a consistent theoretical and experimental framework \cite{Bridgman}.

Dimensions of physical quantities can be formalized as vectors living in a ``physical dimension" vector space. The dimension of a quantity $Q$ in a system with three ``base" dimensions can be expressed as
 \begin{equation}
[Q] \: = \: {\rm M}^\alpha {\rm L}^\beta {\rm T}^\gamma,
 \end{equation}
 where $\rm M,L,T$ correspond to the dimensions of mass, length, and time, respectively. Then, $[Q]$ is a member of a vector space over~$\mathbb{R}$, where vector addition is represented by multiplying $[Q]$ and vector multiplication is represented by raising~$[Q]$ to a power:
  \begin{equation}
[Q_1^a Q_2^b] \: = \: [Q_1]^a [Q_2]^b.
 \end{equation}
The triplet $(\alpha,\beta,\gamma)$ is a particular representation of $[Q]$, dependent on the choice of base dimensions. In a system of different fundamental units (and therefore dimensions), for instance, weight $\rm W$ instead of mass $\rm M$, the dimension of a quantity will remain the same, even though its coordinates will be different (exactly like vectors are independent of the coordinate system we parametrize them in).

The dimensions of a quantity can be used to determine how it transforms under a change of units. For instance, energy has dimension $\rm ML^2 T^{-2}$ and therefore would quadruple if the definition of a meter were to be halved, whereas a dimensionless quantity or ``pure number'' with dimension 1 would not transform. The usual nomenclature of vector spaces can be applied even further: the quantities $Q_i$ are said to have linearly independent dimensions if there is no non-trivial way of constructing a dimensionless quantity out of them, or, equivalently, if 
 \begin{equation}
\left[ \prod_i [Q_i]^{\alpha_i} \right] \: = \: 1
 \end{equation}
 implies $\alpha_i = 0$ for all $i$.

Dimensional analysis fundamentally rests upon the principle that the laws of nature should not depend on our choice of units. This simple idea turns out to have profound implications: it can be used to actually constrain the form of physical relations. This constraint is often expressed in terms of the \emph{Buckingham-$\pi$ theorem} \cite{buckingham}. The simplest statement of this theorem is that a physically meaningful relation between quantities $q_i$ can always be non-dimensionalized, i.e. written in terms of dimensionless quantities $\pi_j$ constructed out of $q_i$. It is not difficult to see why this is the case: since, by definition, only dimensionless quantities are invariant under changes of units, then any invariant law must be constructed from dimensionless quantities only. More specifically, if we have a set of $m$ dimensionful quantities with $n$ linearly independent dimensions, and $k$ dimensionless quantities $c_j$, then any non-dimensionalized law can be written as
 \begin{equation}\label{bpt}
 f(\pi_1, \ldots, \pi_{m-n}, c_1, \ldots c_k) \:  = \: 0.
 \end{equation}
It is not difficult to see that this relation encapsulates all possible relations that remain invariant under a change of units. Rayleigh's method of dimensional analysis (which is what ``dimensional analysis" most often refers to in an undergraduate context), a procedure in which a system is analyzed by considering all the dimensionful independent variables that might influence a dependent variable, is a particular application of the Buckingham-$\pi$ theorem.

The Buckingham-$\pi$ theorem, as expressed through \eref{bpt}, gives us a profound insight on what exactly it means to examine a physical relation in a particular limit. We observe that it is meaningless to talk about the limit of a dimensionful quantity: it only makes sense to consider the limit of the dimensionless quantities $\pi_j$ or $c_j$. This is an important distinction, because once we have decided to include a dimensionful quantity as part of our dimensional analysis, we are considering a different system. Consider, for example, special relativity. Formally (and na\"ively) taking the limit $c\to \infty$ in hopes of recovering classical mechanics can lead to  unexpected results, such as infinite rest mass $mc^2$ for particles. This is because the very concept of ``rest mass'' presupposes the existence of a finite speed limit, which does not vanish simply by taking this limit to infinity.

For the reasons outlined above, studying the correspondence between classical and quantum mechanics is more nuanced than it might originally appear.
A common statement of the correspondence principle is due to Dirac \cite{dirac}, who regarded classical mechanics as the limiting case of quantum mechanics when the (reduced) Planck constant $\hbar \to  0$ (see \cite{klein} for a review). This statement is formally valid for a large scope of applications, but as discussed, may lead to misleading results if applied carelessly. For instance, we might be tempted to write $\Delta x \Delta p >0$ in the classical limit. However, we know that only limits of dimensionless ratios are meaningful. This means that Bohr's statement of the correspondence limit ($n \to \infty$, where $n$ are the quantum numbers of the system in question) is more precise, and the two statements are not necessarily equivalent \cite{liboff}.

In order to illustrate how the correspondence limit of uncertainty bounds can be viewed through the lens of dimensional analysis, we consider a particle with a one-dimensional trajectory. Immediately, from dimensional analysis, we can see that there is no way to impose a limit on $\Delta x$ and $\Delta p$: there are no dimensionless ratios that can be derived from these quantities. We need to introduce at least one additional scale with dimensions of action ($\rm M L^2 T^{-1}$). In quantum mechanics, such a scale already exists: $\hbar$  denotes the scale at which quantum effects are relevant. In the classical ensemble, however, any meaningful uncertainty limit will necessarily be expressed in terms of some physical scale(s) of the system.

We now consider a particle in the quantum realm with characteristic action $A$ (defined as the integral of momentum over length) along with the quantum unit of action $\hbar$. We wish to arrive at a relation between $\Delta x$ and $\Delta p$, and so the dimensionless quantities are given by
\begin{eqnarray}
\pi_1 \: =  \:  \frac{\hbar }{A},  \qquad \pi_2  \: = \:  \frac{\hbar}{\Delta x \Delta p} \; ,
\end{eqnarray}
or, alternatively 
\begin{eqnarray}\label{p12}
\pi_1 \: =  \:  \frac{\hbar }{A},  \qquad \pi_2  \: = \:  \frac{A}{\Delta x \Delta p} \; .
 \end{eqnarray}
Any of these two choices is valid; after all, the Buckingham-$\pi$ theorem does not tell us which choice of dimensionless ratios is more ``physically meaningful''. However, it is clear that if the limit $\hbar/A \to 0$ is taken, the choice (\ref{p12}) contains more information. Alternatively, we may replace   $\pi_1$ by the quantum number $n$, since both tell us how prominent quantum effects are. The uncertainty relation is therefore going to satisfy
\begin{eqnarray}
f \left( \frac{\Delta x \Delta p}{A}, \ \frac{\hbar}{A} \right) \: = \: 0.
 \end{eqnarray}
As a result, the uncertainty equation in the classical limit can be recovered by setting $\hbar/A\to 0$. 

If we instead follow a classical treatment without ever introducing $\hbar$, the uncertainty relation is simply
\begin{eqnarray}
g \left( \frac{\Delta x \Delta p}{A} \right) \: = \: 0,
 \end{eqnarray}
 and by the correspondence principle, we expect
 \begin{eqnarray}
\lim_{\hbar/A \to 0} f \left( \frac{\Delta x \Delta p}{A}, \ \frac{\hbar}{A} \right) \: = \: g \left( \frac{\Delta x \Delta p}{A}  \right)\:.
 \end{eqnarray}
In this limiting case, the minimum uncertainty is necessarily given by $g(\Delta x \Delta p /A) = 0$, and as such, the uncertainty principle is given by
\begin{equation}
\frac{ \Delta x \Delta p}{A} \: > \: c \:,
 \end{equation}
where $c$ is some dimensionless parameter, the value of which cannot be determined via dimensional analysis. However, we observe that na{\"i}vely setting $\hbar = 0$ does not necessarily recover the classical limit $\hbar/A \to 0$; after all, there is no guarantee that the function $f$ is continuous.

For a simple system of a single particle with energy $E$ and mass $m$ in a box of length $L$, we find that the unit of action $A = L \sqrt{2mE}$. However, in general, we expect that the uncertainty principle is going to acquire the following form:
 \begin{equation}
\frac{ \Delta x \Delta p}{A} \: > \:  f(\pi_j) \: ,
 \end{equation}
where the $\pi_j$ are constructed out of the characteristic action scales $A_i$ of the system and $A$ is some weighted average of the $A_i$. 
This is as far as dimensional analysis can take us. We cannot hope of arriving at the precise form of $A$. Importantly, the above discussion presupposes that the notion of ``uncertainty'' can exist in the classical realm. While a truly random variable cannot exist in classical mechanics, chance may be introduced to classical mechanics via \emph{ensembles}, as discussed in the following section. 

\section{Classical probability and uncertainty}

At first glance, it may seem that there is no room for probability in classical mechanics. Laplace envisioned a being, ``Laplace's demon'', which ``at a certain moment would know all forces that set nature in motion''. He argued that for such a being, ``nothing would be uncertain and the future just like the past would be present before its eyes'' \cite{laplace}. However, even in such a perfectly deterministic world, there is room for the notion of probability. Indeed, the proper classical limit of quantum mechanics (and the only avenue by which we can describe ``classical probabilty'' in any meaningful sense) is classical statistical mechanics rather than classical mechanics \cite{ballentine, huang}, which necessitates the use of the statistical ensemble if we are to extend the notion of uncertainty into the classical realm.

If a single-particle system is viewed as a member of an \textit{ensemble} of bound-state particles, the position of the particle is no longer a trajectory, but rather a distribution. We may view this distribution either in the frequentist sense of a random selection of one of the members of the ensemble, or in the Bayesian sense of ``parametrizing our ignorance'' of the particle's initial conditions. 
The distribution for the position of a particle in a classical ensemble is derived by observing that the particle spends a time $dt = dx/|v(x)|$ within the interval $dx$, where $v(x)$ is its speed. Assuming we observe the particle at some unspecified time, the classical probability density for position measurements of a particle uniformly selected from such an ensemble is defined as
\begin{equation} \label{density}
 \rho(x) \: = \: \frac{N}{\sqrt{2m[ E-V(x)]}},
\end{equation}
where $N$ is a normalization constant, $E$ is the energy of the bound state and $V(x)$ the potential function. This confirms our intuition that the particle is found more often in regions where its speed is low. For periodic systems, $N =2/\tau$ where $\tau$ is the period.

From this definition of classical probability density, it follows that the average of a classical operator $A$ can be defined as an average over one period $\tau$:
\begin{equation} \label{tav}
    \langle {A} \rangle_{\mathrm{CL}} \:  = \:  \frac{\int_{0}^{\tau}A(t) dt}{\int_{0}^{\tau} \: dt}.
\end{equation}
Furthermore, it is possible to straightforwardly calculate the variance of classical operators using the usual definition $\Delta A^2 = \langle A^2 \rangle - \langle A \rangle^2$.

Having put all preliminaries in place, we may turn our attention to the correspondence between the classical and quantum uncertainty in a few select systems of interest. Single-particle systems have been examined thoroughly in the literature, and from a dimensional analysis perspective, two-body systems may be more robust. However, in order to set the scene and allow for a direct comparison with coupled systems, we will cite here the one-particle results for the cases of harmonic oscillator and infinite square well. As discussed in the previous section, it is illuminating to view uncertainty bounds through a non-dimensionalized system of equations, which motivates us to define the dimensionless position and momentum variables as follows
\begin{equation} \label{scaled_xp}
    \overline{x}_i=\frac{x_i}{A_i}, \quad \quad \quad \overline{p}_i=\frac{p_i}{m_i \omega_i A_i},
\end{equation}
where $A_i$ is the initial amplitude of the system, $m_i$ its mass and $\omega_i$ is the frequency of oscillation. 

With this notation, the uncertainty product for position and momentum in the classical harmonic oscillator case is found to be 
\begin{equation} \label{uncoupled}
    \Delta \overline{x}\Delta \overline{p} \:=\: \frac{1}{2}.
\end{equation}
Similarly, for a single particle confined to move under a one-dimensional infinite square well potential, the uncertainty product is
\begin{equation} \label{3case}
    \Delta \overline{x}  \Delta \overline{p} \:=\: \frac{1}{\sqrt{12}}.
\end{equation}
This coincides with the quantum mechanical result in the limit of large principal quantum number $n$. For a full derivation of these uncertainty results, the reader is referred to \cite{usha} and \cite{R6}. 

For single-particle systems such as the harmonic oscillator and the infinite square well, we note that the uncertainty bound reduces to a pure number in the classical limit. This is to be expected, since there are just enough dimensionful parameters in order to define the dimensionless analogues of position and momentum. In the presence of additional dimensionful parameters, however, we expect the uncertainty bounds to depend on residual dimensionless variables even in the limit of large quantum numbers. Coupled systems feature more dimensionful parameters, and for this reason we turn our attention to them in the next section. 

In concluding this section, we must stress that recasting quantum mechanics in the classical phase space (effectively deducing its postulates on statistical grounds) is a much more involved procedure than assigning a classical distribution to a particle ensemble as outlined above. This procedure usually involves ``geometrizing" quantum mechanics by constructing the phase space of a classical system, which is then endowed with a probability measure.
Riccia and Wiener~\cite{riccia}, for instance, use stochastic integrals in order to motivate Born's rule, while Kibble~\cite{kibble} employs symplectic manifolds with a complex structure in order to recover quantum dynamics. Another geometric approach by Heslot \cite{heslot} reveals $\hbar$ to be the curvature of the space of quantum states, indicating once again that there is a qualitative difference between the classical and quantum realms; the former features a state space which has a natural unit of distance, while the latter does not. The field of quantum information~\cite{weedbrook, luo} features powerful and sophisticated techniques that make the quantum-classical correspondence much more manifest. These approaches to the relation between classical and quantum mechanics are quite involved, and beyond the scope of the present work. Nonetheless, our simple approach still highlights the often-overlooked role that dimensional analysis plays in both classical and quantum uncertainty and the classical limit of quantum mechanics.

\section{Two coupled harmonic oscillators with equal masses}\label{equal_masses}
We consider a simple coupled system consisting of two harmonic oscillators with spring constant $k$ and equal masses $m$, coupled via a potential of the form $V(x_1, x_2)~=~\frac{1}{2}k'(x_1\:-\:x_2)^2$, where $k \neq k'$ in general, and $x_1$ and $x_2$ denote the displacements of the individual oscillators. For similar treatments of this set-up the reader is referred to \cite{rachel, amore, marsiglia}.

\subsection{Classical oscillators} \label{CO_m}
We model the classical system with two blocks constrained to move in a one-dimensional frictionless surface. Each block is attached to an outer stationary wall by means of a spring with force constant $k$. The inner spring has force constant $k'$. It is assumed that all springs assume their natural length when the system is at rest.

This 1D problem can be easily approached utilizing the Lagrangian formalism and is part of the repertoire of any undergraduate dynamics course. The Lagrangian of the system reads
\begin{equation} \label{lagr}
    \mathcal{L}\:=\: \frac{1}{2} m\dot{x}_1^2\:+\: \frac{1}{2}m \dot{x}_2^2 \:-\:\frac{1}{2}k(x_1^2\:+\:x_2^2)\:-\:\frac{1}{2}k'(x_1\:-\:x_2)^2.
\end{equation}
To decouple (\ref{lagr}) and obtain the equations of motions, we perform a change of variables in favour of the normal mode coordinates $x_{\mathrm{c},\mathrm{r}}$ defined as linear combinations of the position variables $x_{1,2}$, i.e.
\begin{equation}\label{normal}
    x_{\mathrm{c}} \:=\: \frac{x_1 \:+\: x_2}{\sqrt{2}}, \quad \quad \quad x_{\mathrm{r}} \:=\: \frac{x_1 \:-\: x_2}{\sqrt{2}},
\end{equation}
which are the centre of mass and relative distance respectively. 

The characteristic frequencies of the two normal modes of oscillation are given by $\omega_{\mathrm{c}}^2\:=\:k/m$ and $\omega_{\mathrm{r}}^2 \:=\: (k\:+\:2k')/m$, corresponding to in phase and out of phase oscillations respectively.\par
The time averages of the normal modes coordinates for position and momentum are then found through \eref{tav} over a period $\tau_i \:=\: 2 \pi /\omega_i$ with $i\: = \:  \mathrm{c},\mathrm{r}\: $:
\begin{eqnarray}
    \langle x_i\rangle \:&=\: 0, \qquad \langle x_i^2\rangle \:&=\: \frac{A_i^2}{2},  \label{meanq1} \\
    \langle p_i\rangle \:&=\: 0 ,\qquad \langle p_i^2\rangle \:&=\: \frac{1}{2}m^2 \omega_i^2A_i^2. \label{meanq3}
\end{eqnarray}
It is now straightforward to change back to the position variables $x_{1,2}$ in order to find the uncertainty relation in its conventional formulation as the product of the uncertainties $\Delta x$ and $\Delta p$.

The time averages for the position variable $x_1$ can be readily obtained via of (\ref{meanq1}) and (\ref{meanq3}):
\begin{eqnarray}
    \langle x_1 \rangle \:&=\: \frac{\sqrt{2}}{2} \langle x_{\mathrm{c}} \:+\:  x_{\mathrm{r}} \rangle \:=\: 0, \\
    \langle x_1^2 \rangle \:&=\: \frac{1}{2} \langle ( x_{\mathrm{c}} \:+\:  x_{\mathrm{r}})^2 \rangle \:=\: \frac{1}{4}(A_{\mathrm{c}}^2 \:+\: A_{\mathrm{r}}^2).
\end{eqnarray}
This yields the following expression for the variance of $x_1$:
\begin{equation}
    \Delta x_1 \:=\: \sqrt{\langle x_1^2 \rangle-\langle x_1 \rangle^2}=\frac{\sqrt{A_{\mathrm{c}}^2+A_{\mathrm{r}}^2}}{2}.
\end{equation}
Using the definition, the time averages and variance of the momenta  are found to be
\begin{eqnarray}
    \langle p_1 \rangle \:&=\: 0,\qquad
    \langle p_1^2 \rangle \:=\: \frac{1}{4}m^2(A_{\mathrm{c}}^2\omega_{\mathrm{c}}^2 \:+\: A_{\mathrm{r}}^2\omega_{\mathrm{r}}^2),\\ 
    \Delta p_1 \:&=\:  m\frac{\sqrt{A_{\mathrm{c}}^2\omega_{\mathrm{c}}^2 \:+\: A_{\mathrm{r}}^2\omega_{\mathrm{r}}^2}}{2}.
\end{eqnarray}
Repeating the previous steps for the second block yields identical results. The product of the variances is therefore
\begin{equation} \label{x11}
    \Delta x_1\Delta p_1 \:=\:  m\frac{\sqrt{A_{\mathrm{c}}^2 \:+\: A_{\mathrm{r}}^2}\sqrt{A_{\mathrm{c}}^2\omega_{\mathrm{c}}^2 \:+\: A_{\mathrm{r}}^2\omega_{\mathrm{r}}^2}}{4},
\end{equation}
and similarly for $\Delta x_2\Delta p_2$. Using the scaled canonical variables defined in (\ref{scaled_xp}), (\ref{x11}) can be written as
\begin{equation} \label{x12}
    \Delta \overline{x}_1\Delta \overline{p}_1 \:=\: \frac{1}{4}\sqrt{1 \:+\: \frac{A_{\mathrm{c}}^2}{A_{\mathrm{r}}^2}}\sqrt{1 \:+\: \frac{A_{\mathrm{c}}^2\omega_{\mathrm{c}}^2}{A_{\mathrm{r}}^2\omega_{\mathrm{r}}^2}}.
\end{equation}
where the variables have been scaled by $A_{\mathrm{r}}$ and $m A_{\mathrm{r}} \omega_{\mathrm{r}}$ respectively.
Since the system is symmetric under particle exchange, the expression for $\Delta x_2 \Delta p_2$ can be easily deduced from that of $\Delta x_1 \Delta p_1$ by swapping indices $1$ and $2$.
We note that this expression, which represents the uncertainty relation in terms of dimensionless position and momentum for a classical ensemble, comprises of two dimensionless ratios involving the amplitudes~$A_{\mathrm{c,r}}$ and frequencies $\omega_{\mathrm{c,r}}$ respectively, as expected.

We may compare the uncertainty product of this system to that of the 1D uncoupled case of (\ref{uncoupled}) by setting $A_{\mathrm{c}} \:=\: A_{\mathrm{r}}$ and $\omega_{\mathrm{c}} \:=\: \omega_{\mathrm{r}}$. As such, we observe that the effect of the coupling is to add another dimensionful degree of freedom to the system, which results in the possibility for another dimensionless ratio to appear in the non-dimensionalized expression for the uncertainty product.

\subsection{Quantum oscillators} \label{52}
The quantum mechanical analogue of the system we have analyzed so far is that of two distinguishable coupled one-dimensional harmonic oscillators. We treat the oscillators as distinguishable to allow for a direct comparison to the uncertainty product of \eref{x12} in the previous section. The system is then described by the following Hamiltonian:
\begin{equation}
    \hat{H} \:=\: \frac{\hat{p}_1^2 \:+\: \hat{p}_2^2}{2m} \:+\: \frac{1}{2}k(x_1^2 \:+\: x_2^2) \:+\: \frac{1}{2}k'(x_1 \:-\: x_2)^2.
\end{equation}
Using a variation of the Jacobi coordinates allows to reduce the complexity of the problem by transforming it to what is essentially a one-dimensional problem in terms of the centre of mass and relative coordinate of \eref{normal} \cite{greiner}.

After the change of variables, the Hamiltonian becomes separable:
\begin{eqnarray}
  \hat{H} \:&=\: \frac{\hat{p}_{\mathrm{c}}^2}{2m} \:+\: \frac{1}{2}k x_{\mathrm{c}}^2 \:+\: \frac{\hat{p}_{\mathrm{r}}^2}{2m} \:+\: \frac{1}{2}(k \:+\: 2k') x_{\mathrm{r}}^2  \\
    \:&=\: \hat{H}_{\mathrm{c}} \:+\: \hat{H}_{\mathrm{r}}.
\end{eqnarray}
It is therefore possible to determine the evolution of the system by viewing it as two single-particle harmonic oscillators with angular frequencies given by $ \omega_{\mathrm{c}}^2 \:=\: k/m$ and $ \omega_{\mathrm{r}}^2 \:=\: (k+2k')/m$.
Since the oscillators are treated as distinguishable, the total wavefunction of the system is the product of the single-particle wavefunctions
\begin{equation}\label{wave}
    \Psi(x_{\mathrm{c}},x_{\mathrm{r}}) \:=\:  \phi_{n_{\mathrm{c}}}(x_{\mathrm{c}}) \phi_{n_{\mathrm{r}}}(x_{\mathrm{r}}),
\end{equation}
where 
\begin{equation} \label{ho}
    \phi_n(x) \:=\: \left(\frac{\sqrt{\pi}}{x_0 2^n n!}\right)^{1/2}H_n\left(\frac{x}{x_0}\right)\exp\left(-\frac{x^2}{2x_0^2}\right),
\end{equation}
with $x_0 \:=\: \sqrt{\frac{\hbar}{m \omega}}$, and $n_{\mathrm{r}}$ and $n_{\mathrm{c}}$ are the principal quantum numbers associated with the wavefunctions $\phi(x_{\mathrm{r}})$ and $\phi(x_{\mathrm{c}})$ respectively.
\begin{figure}
\centering
\subfloat[\label{fig:sfig13}]{%
  \includegraphics[width=0.3\columnwidth]{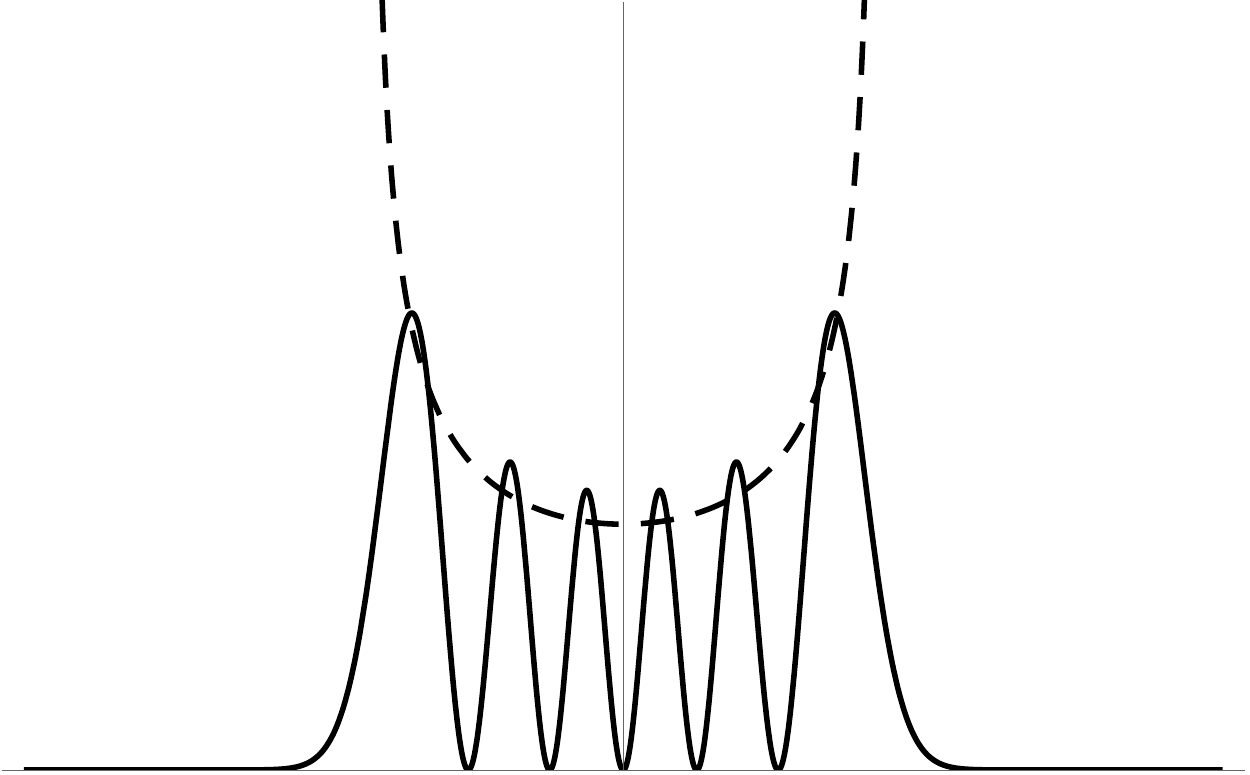}%
}
\subfloat[\label{fig:sfig21}]{%
  \includegraphics[width=0.3\columnwidth]{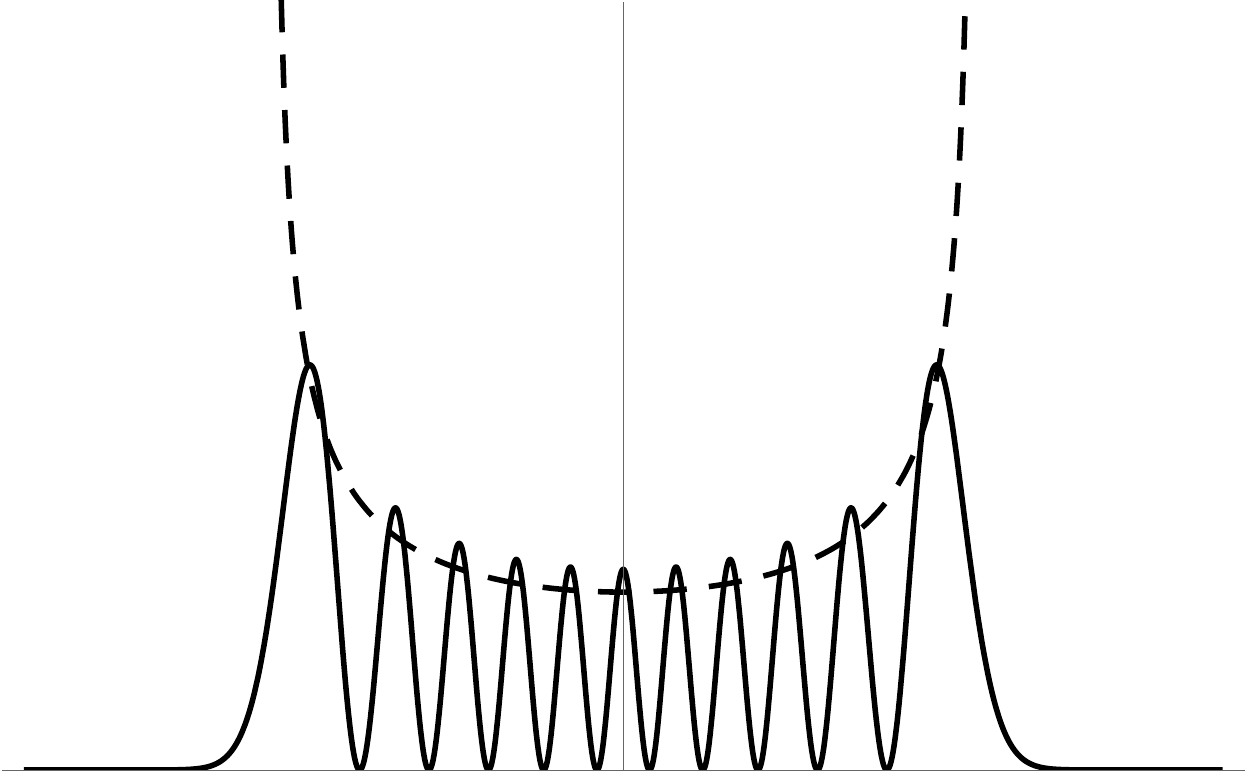}%
}
\subfloat[\label{fig:sfig22}]{%
  \includegraphics[width=0.3\columnwidth]{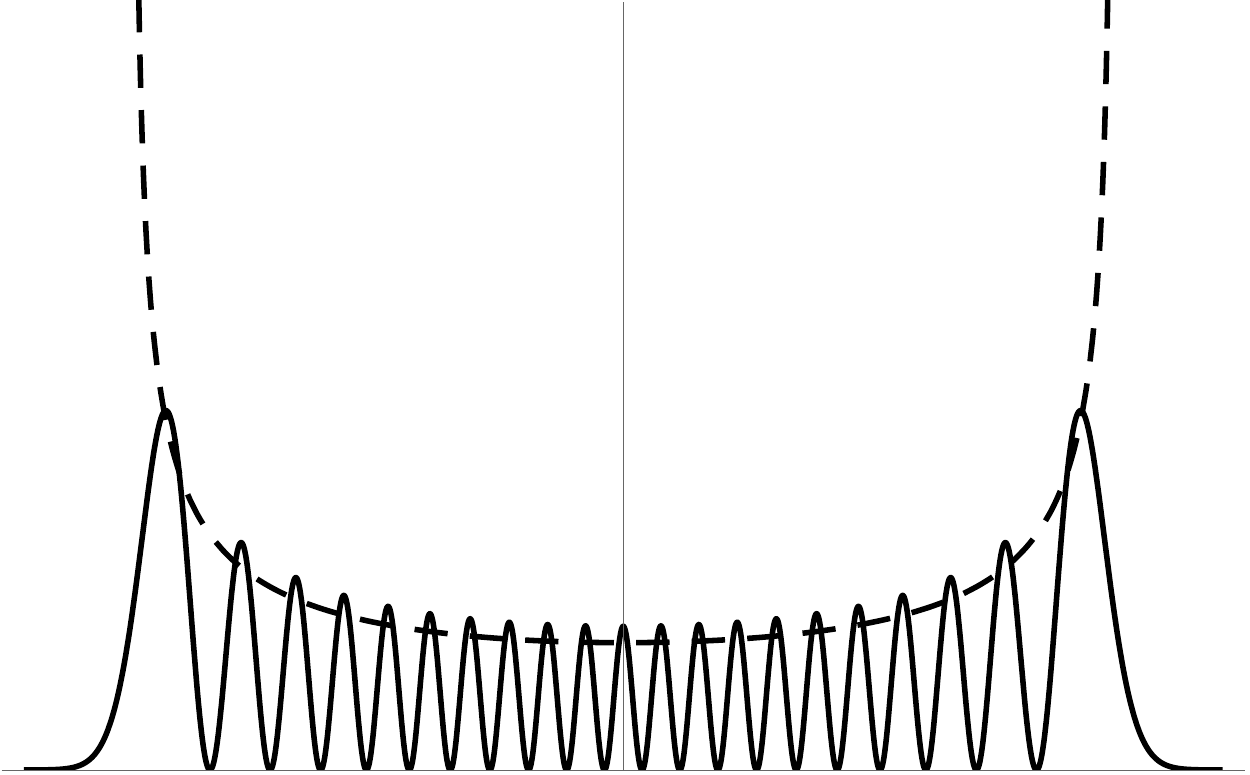}%
}
\caption{Uncoupled classical (dashed line) and quantum (solid line) probability densities for the harmonic oscillator as functions of the centre of mass coordinate $x_{\mathrm{c}}$ for (a) $n_{\mathrm{c}}=5$, (b) $n_{\mathrm{c}}=10$ and (c) $n_{\mathrm{c}}= 20$.}\label{fig_two}
\end{figure}
\begin{figure}
\centering
\subfloat[\label{fig:sfig14}]{%
  \includegraphics[width=0.3\columnwidth]{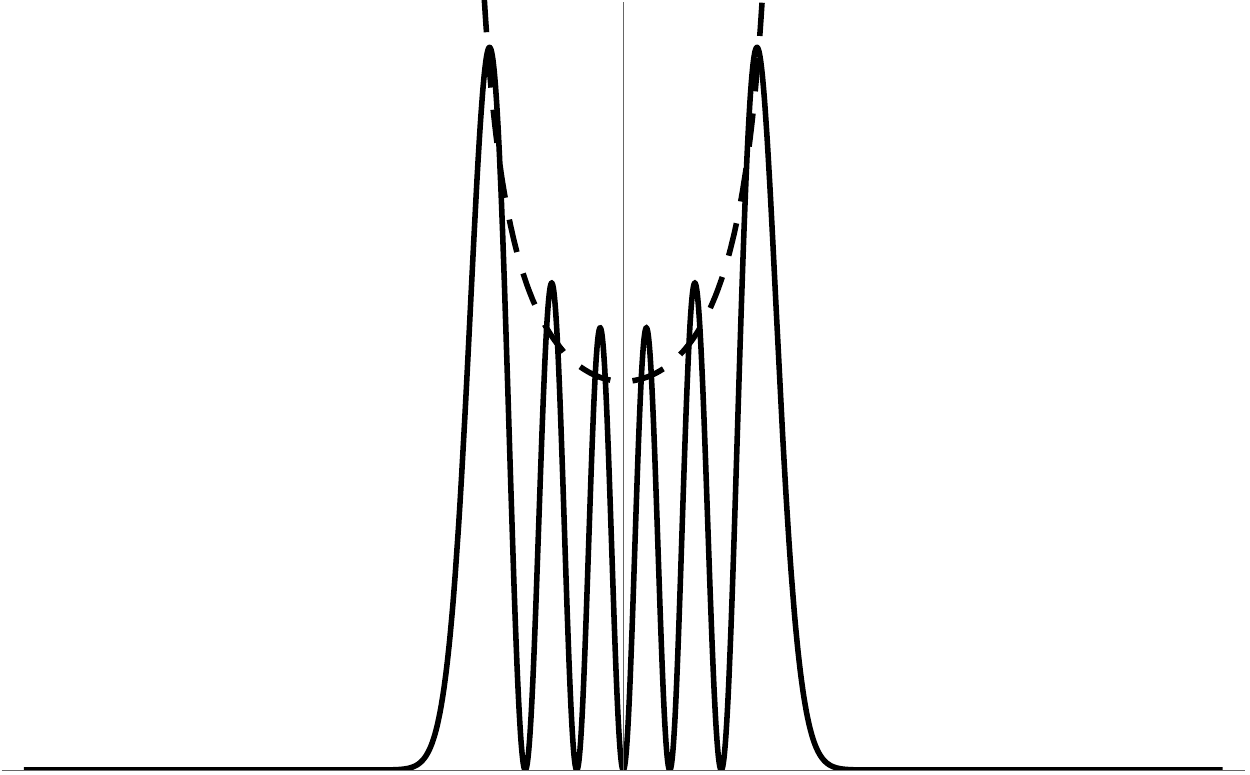}%
}
\subfloat[\label{fig:sfig24}]{%
  \includegraphics[width=0.3\columnwidth]{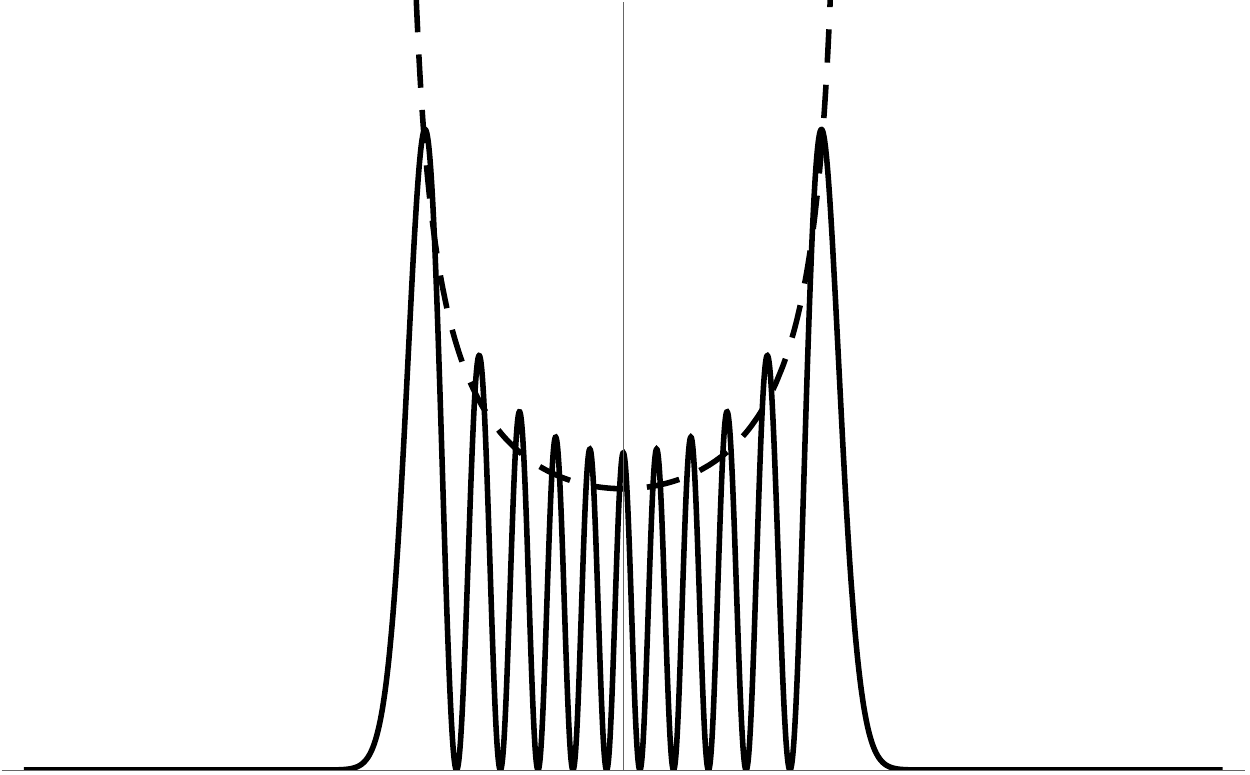}%
}
\subfloat[\label{fig:sfig30}]{%
  \includegraphics[width=0.3\columnwidth]{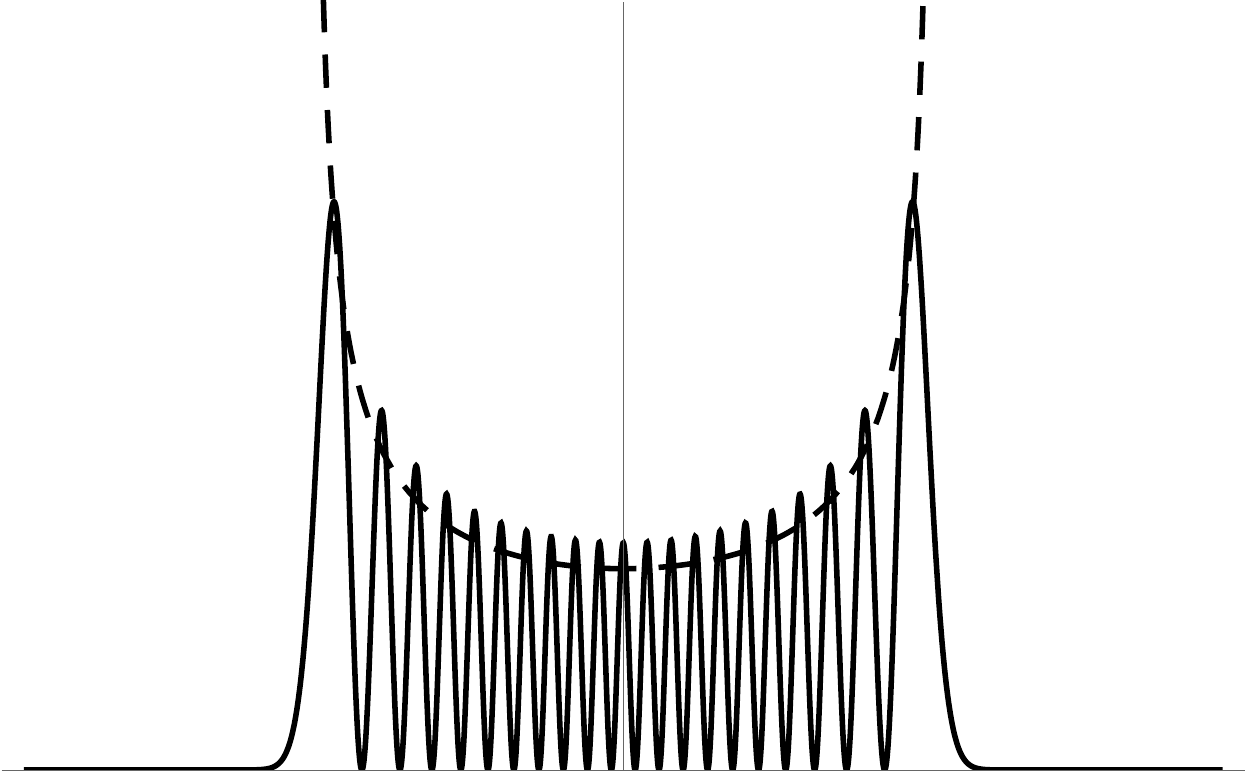}%
}
\caption{Uncoupled classical (dashed line) and quantum (solid line)  probability densities for the harmonic oscillator as functions of the relative  coordinate $x_{\mathrm{r}}$ for (a) $n_{\mathrm{r}}=5$, (b) $n_{\mathrm{r}}=10$ and (c) $n_{\mathrm{r}}= 20$.}\label{fig_three}
\end{figure}

Using the definition for quantum mechanical expectation values and making use of the orthogonality properties of the Hermite polynomials, one can calculate the expectation values for $x_{\mathrm{r}}$ and $x_{\mathrm{c}}$ and the respective momentum operators for the general wavefunction defined in \eref{wave} as
\begin{eqnarray} 
    \langle x_i\rangle \:&=\: 0, \quad \quad \quad \langle x_i^2\rangle \:=\: \frac{(2n_i \:+\: 1)\hbar}{2m\omega_i},\label{align1}\\
    \langle p_i\rangle \:&=\: 0, \quad \quad \quad \langle p_i^2\rangle \:=\: (2n_i \:+\: 1) \frac{m\hbar\omega_i}{2}, \label{align2}
\end{eqnarray}
where $i \: =  \:  \mathrm{c}, \mathrm{r} $.
We can now transform back to the original spatial variables~$x_1$ and~$x_2$ and obtain the expectation value for positions and momenta using the results~\eref{align1} and~\eref{align2} above:
\begin{eqnarray}
\fl   \langle x_1 \rangle \:&=\: 0, \quad \quad \quad
    \langle x_1^2 \rangle \:&=\: \frac{1}{2}\left[\frac{(2n_{\mathrm{r}} \:+\: 1)\hbar}{2m\omega_{\mathrm{r}}} \:+\: \frac{(2n_{\mathrm{c}} \:+\: 1)\hbar}{2m\omega_{\mathrm{c}}}\right],\\
\fl \langle p_1 \rangle \:&=\: 0, \quad \quad \quad
    \langle p_1^2 \rangle \:&=\: \frac{1}{2}\left[\frac{m \hbar \omega_r(2n_{\mathrm{r}} \:+\: 1)}{2}+\frac{m\hbar \omega_{\mathrm{c}}(2n_{\mathrm{c}} \:+\: 1)}{2}\right],
\end{eqnarray}
and similarly for the second oscillator.
Hence, the quantum uncertainties in position and momentum read
\begin{figure}
\centering
\subfloat[\label{fig:sfig11}]{%
  \includegraphics[width=0.3\columnwidth]{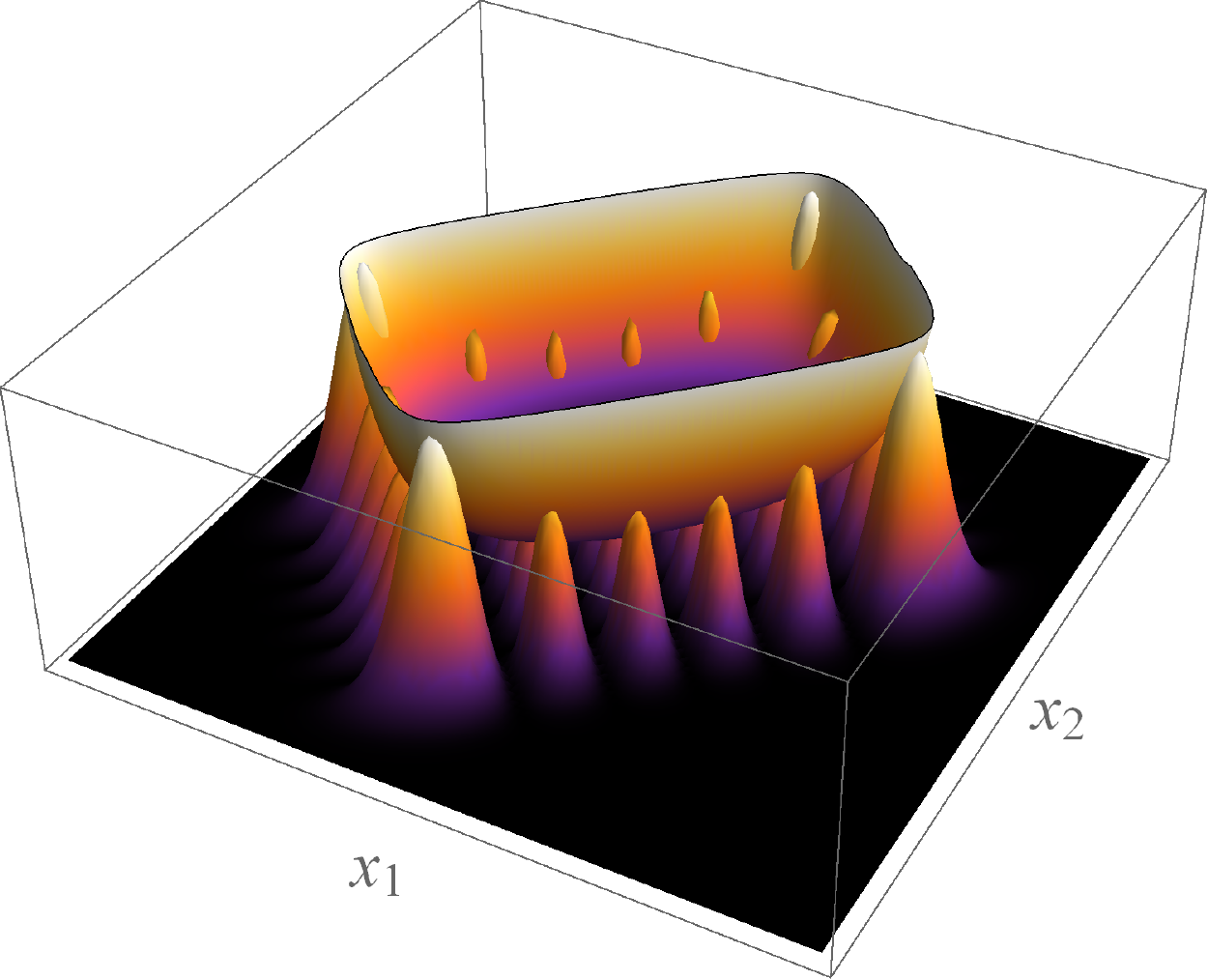}%
}
\subfloat[\label{fig:sfig26}]{%
  \includegraphics[width=0.3\columnwidth]{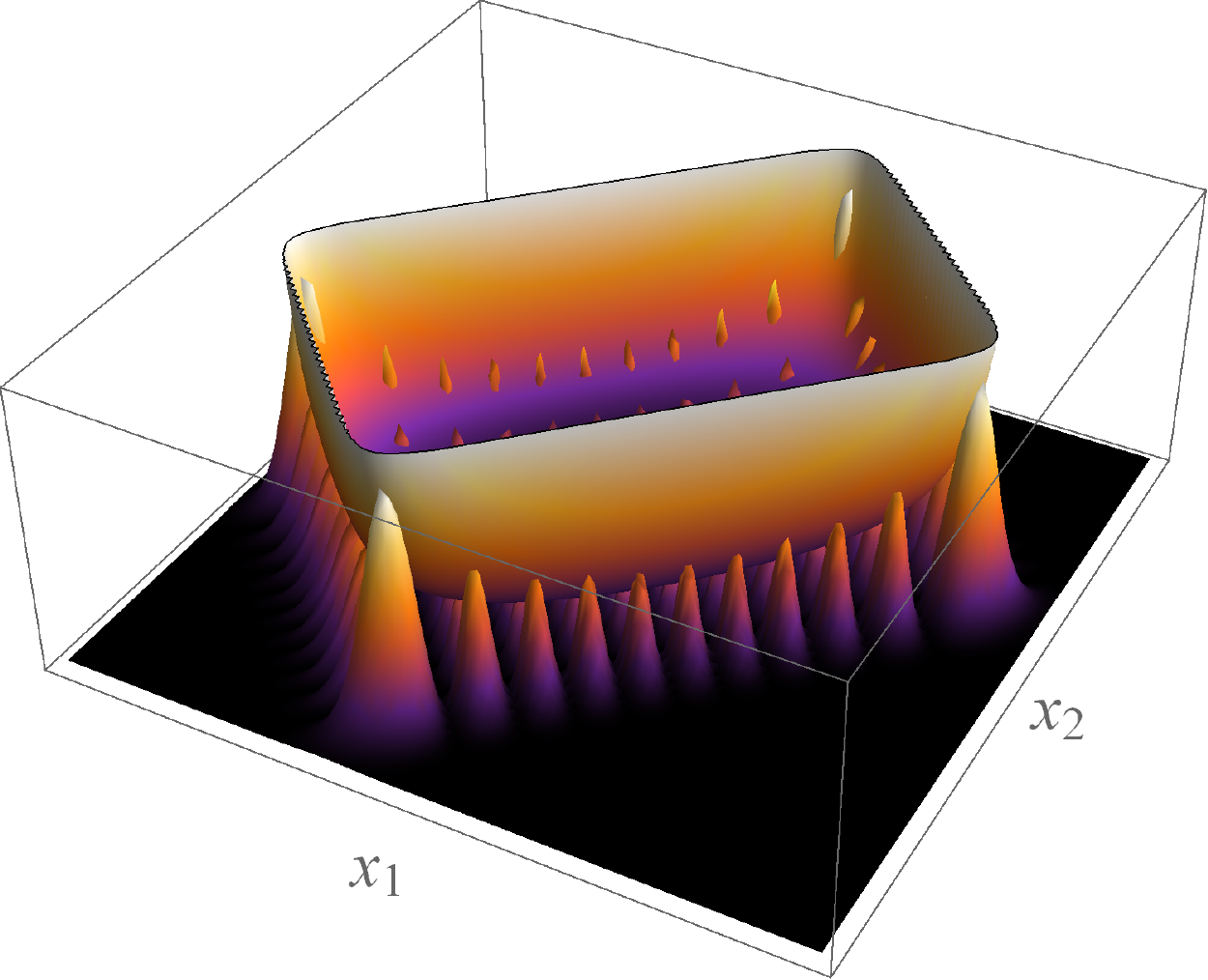}%
}
\subfloat[\label{fig:sfig31}]{%
  \includegraphics[width=0.3\columnwidth]{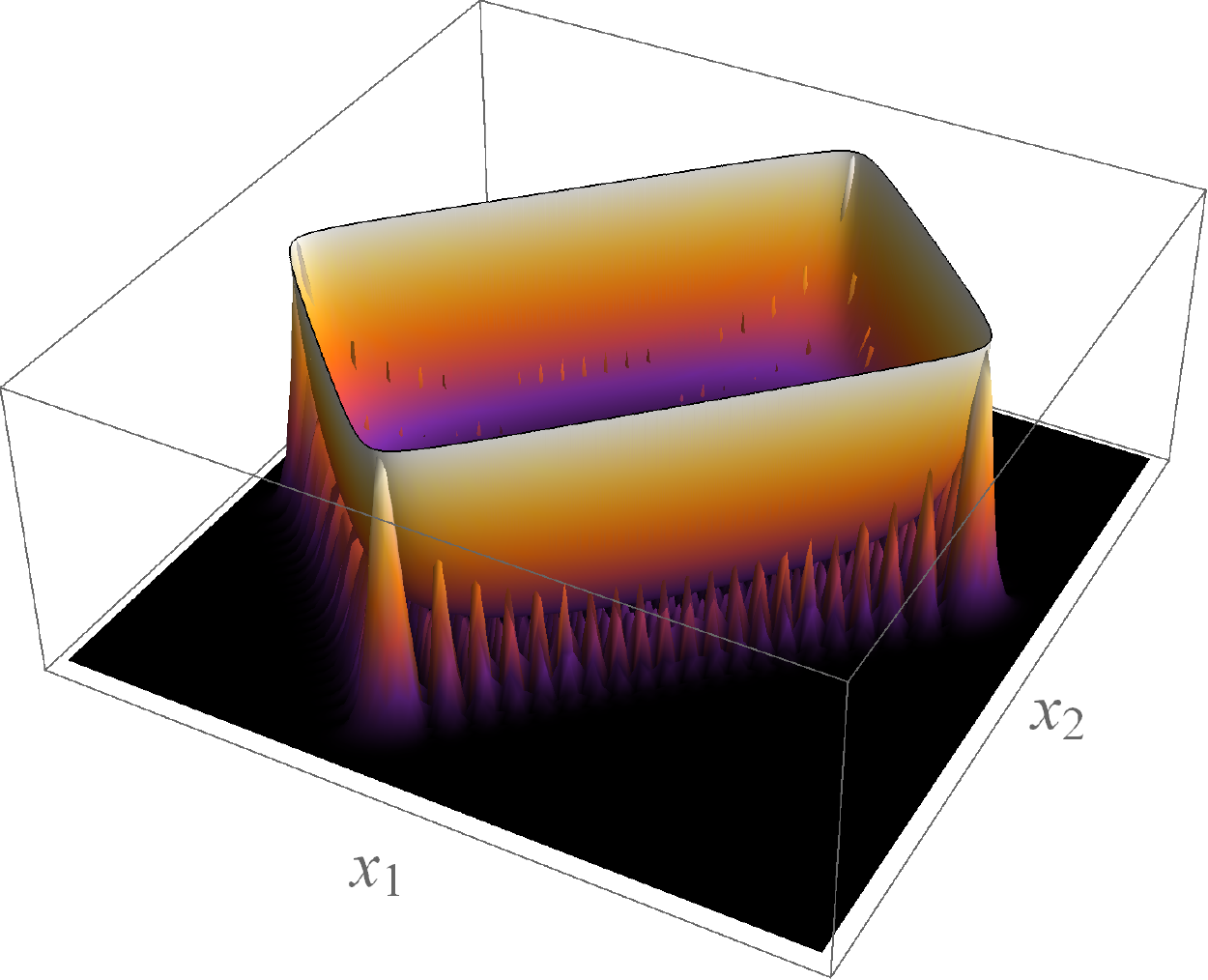}%
}
\caption{Classical $\rho_{\mathrm{CL}}(x_1, x_2)$ and quantum $\rho_{\mathrm{QM}} = |\psi_n(x_1, x_2)|^2$ probability densities for coupled harmonic oscillators with equal masses vs the spatial variables $x_{1}$ and $x_2$ for (a) $n=5$, (b) $n=10$, (c) $n=20$.}\label{fig_one}
\end{figure}

\begin{eqnarray}
    \Delta x_1 \:&=\: \frac{1}{2} \sqrt{\frac{(2n_{\mathrm{r}} \:+\: 1)\hbar}{m\omega_{\mathrm{r}}} \:+\: \frac{(2n_{\mathrm{c}} \:+\: 1)\hbar}{m\omega_{\mathrm{c}}}} \\ \:&=\: \frac{1}{2} \sqrt{A_{n_{\mathrm{c}}}^2 \:+\: A_{n_{\mathrm{r}}}^2},
\end{eqnarray}
\begin{eqnarray}
     \Delta p_1 \:&=\: \frac{1}{2} \sqrt{m \hbar \omega_{\mathrm{r}}(2n_{\mathrm{r}} \:+\: 1)\:+\:m\hbar \omega_{\mathrm{c}}(2n_{\mathrm{c}} \:+\: 1)}  \\
    \:&=\: \frac{m}{2} \sqrt{\omega_{\mathrm{c}}^2 A_{n_{\mathrm{c}}}^2 \:+\: \omega_{\mathrm{r}}^2 A_{n_{\mathrm{r}}}^2},
\end{eqnarray}
respectively, where $A_{n_{\mathrm{c},\mathrm{r}}}$ are the classical turning points associated with the energy $E_{n_{\mathrm{c},\mathrm{r}}}$ of an harmonic oscillator with mass $m$ and frequency~$\omega_{\mathrm{c},\mathrm{r}}$:
\begin{equation} \label{amp}
    A_{n_{\mathrm{c},\mathrm{r}}}\:=\: \sqrt{\frac{2 E_{n_{\mathrm{c},\mathrm{r}}}}{m \omega_{\mathrm{c},\mathrm{r}}^2}} \:=\: \sqrt{\frac{(2n_{\mathrm{c},\mathrm{r}} \:+\: 1)\hbar}{m \omega_{\mathrm{c},\mathrm{r}} n_{\mathrm{c},\mathrm{r}}}}.
\end{equation} 
In terms of the dimensionless variables $\overline{x}_i$ and $\overline{p}_i$ of \eref{scaled_xp} scaled by means of the amplitude $A_{n_{\mathrm{c},\mathrm{r}}}$, the uncertainty product can be written as
\begin{equation} \label{unc}
    \Delta \overline{x}_1 \Delta \overline{p}_1 \:=\: \frac{1}{4}\sqrt{1 \:+\: \frac{A_{n_{\mathrm{c}}}^2}{A_{n_{\mathrm{r}}}^2}}\sqrt{1 \:+\: \frac{A_{n_{\mathrm{c}}}^2\omega_{\mathrm{c}}^2}{A_{n_{\mathrm{r}}}^2\omega_{\mathrm{r}}^2}},
\end{equation}
or explicitly in terms of the principal quantum numbers $n_{\mathrm{c},\mathrm{r}}$ and harmonic oscillator
frequencies $\omega_{\mathrm{c},\mathrm{r}}$ as
\begin{equation}
    \Delta \overline{x}_1 \Delta \overline{p}_1 \:=\: \frac{1}{4}\sqrt{1 \:+\: \frac{(2n_{\mathrm{c}} \:+\: 1)\omega_{\mathrm{r}}}{(2n_{\mathrm{r}} \:+\: 1)\omega_{\mathrm{c}}}}\sqrt{1 \:+\: \frac{(2n_{\mathrm{c}} \:+\: 1)\omega_{\mathrm{c}}}{(2n_{\mathrm{r}} \:+\: 1)\omega_{\mathrm{r}}}}.
\end{equation}
We note that the last expression only involves dimensionless ratios of the two frequencies of oscillation and the characteristic principal quantum numbers. 
It is suggestive of the correspondence principle that \eref{unc} is in complete agreement with the classical uncertainty of \eref{x12}, and all observations on how the couplings affects the classical uncertainty product also apply in this case. 

To offer a graphical comparison between the classical and quantum system, let~$\rho_{\mathrm{CL}}(x_{\mathrm{c}, \mathrm{r}})$ be the classical probability density defined by
\begin{equation}\label{rhoHO}
    \rho_{\mathrm{CL}}(x) \:=\: \frac{1}{\pi}\frac{1}{\sqrt{A_{n}^2 \:-\: x^2}},
\end{equation}
and valid in the range $x \in\left(-A_{n}, A_{n}\right)$ where $A_{n}$ is defined in (\ref{amp}) \cite{usha, R6}. The usual representation of 1D quantum and classical probability densities can be found in the literature (see \cite{R6}); as such, we illustrate the coupled case. Figure \ref{fig_two} shows the quantum probability density, $\rho_{\mathrm{QM}}(x_{\mathrm{c}})$, along with the classical analogue $\rho_{\mathrm{CL}}(x_{\mathrm{c}})$ in terms of the centre of mass coordinate for various values of the principal quantum number $n_{\mathrm{c}}$. Figure \ref{fig_three} shows the probability densities for the same values of the principal quantum number $n_{\mathrm{r}}$ as a function of the relative coordinate $x_{\mathrm{r}}$ instead.

Finally, let the total classical probability density be given by the product of the ones associated to the individual modes of oscillation $x_{\mathrm{c},\mathrm{r}}$ as
\begin{equation}
    \rho_{\mathrm{CL}}(x_1, x_2) \:=\:  \rho_{\mathrm{CL}}(x_{\mathrm{c}})  \rho_{\mathrm{CL}}(x_{\mathrm{r}}).
\end{equation}
Figure \ref{fig_one} offers a comparison between the quantum probability density, $\rho_{\mathrm{QM}}$ 
 and the classical counterpart $ \rho_{\mathrm{CL}}(x_1, x_2)$ which are plotted as functions of the original spatial variables of the problem for value of $n$ equal to 5, 10 and 20.  Noticeably, the shape of the three dimensional classical probability density is that of a rectangular well with soft edges. This is the natural and intuitive extension of the 1D distribution shown in Figures \ref{fig_two} and \ref{fig_three}, which can be retrieved by slicing the three dimensional distributions across lines of $x_1 \:=\: 0 $ and $x_2 \:=\: 0$.

\section{Two coupled harmonic oscillators with different masses} \label{different_m}
Let us now consider a slight variation of the problem analyzed in the previous section for which the oscillators have masses $m_{1}$ and $m_2$ with  $m_1 \neq m_2$ and the potential function is $V(x_1, x_2)\:=\:\frac{1}{2}k(x_1 \:-\: x_2)^2$. We expect that the results in this case will have a similar form, even though the dimensionless ratios will differ. For similar treatments of this set-up the reader is referred to \cite{rachel, amore}.

\subsection{Classical oscillators}\label{classical}
Similarly to Section \ref{CO_m}, the classical system consists of two blocks connected to three springs. The motion of the blocks is described by the following Lagrangian
\begin{equation} \label{lagrangian_cl}
\fl    \mathcal{L} \:=\: \frac{1}{2} m_1\dot{x}_1^2 \:+\: \frac{1}{2}m_2 \dot{x}_2^2 \:-\: \frac{1}{2}\omega^2(m_1 x_1^2 \:+\:  m_2 x_2^2) \:-\: \frac{1}{2}k(x_1 \:-\: x_2)^2,
\end{equation}
where we assume the angular frequency $\omega$ to be the same for both oscillators.
The equations of motion obtained from Lagrange's equations can be decoupled by performing a variable transformation to the Jacobi coordinates
\begin{equation} \label{xrc} 
    x_{\mathrm{r}} \:=\: \frac{x_1 \:-\: x_2}{\sqrt{2}},\quad \quad \quad x_{\mathrm{c}} \:=\: \frac{m_1 x_1 \:+\: m_2 x_2}{M\sqrt{2}},
\end{equation}
where $M=(m_1 +m_2)/2$. 
Proceeding with this transformation, (\ref{lagrangian_cl}) can be rewritten as
\begin{equation}
    \mathcal{L} \:=\: \frac{1}{2} M\dot{x}_{\mathrm{c}}^2+ \frac{1}{2}\mu \dot{x}_\mathrm{r}^2 \:-\: \frac{1}{2}M\omega_{\mathrm{c}}^2 x_{\mathrm{c}}^2 \:-\: \frac{1}{2}\mu\omega_{\mathrm{r}}^2x_{\mathrm{r}}^2,
\end{equation}
where $\mu \:=\: (m_1 m_2)/M$ is the reduced mass and $\omega_{\mathrm{c}}^2=\omega^2$, $\omega_{\mathrm{r}}^2=\omega^2\:+\:2k/\mu$ are the two harmonic oscillator frequencies.
The time averages of the relative and centre of mass coordinate are given once again by \eref{meanq1}. 
We may then easily transform back to the spatial coordinates $x_1$ and $x_2$ by combining the expressions for $x_{\mathrm{r}}$ and $x_{\mathrm{c}}$ from~\eref{xrc}, i.e.
\begin{equation}
    x_1 \:=\: \frac{1}{\sqrt{2}}\left(x_{\mathrm{c}} \:+\: \frac{m_2}{M}x_{\mathrm{r}}\right), \label{origin1}\quad \quad 
    x_2 \:=\: \frac{1}{\sqrt{2}}\left(x_{\mathrm{c}} \:-\: \frac{m_1}{M}x_{\mathrm{r}}\right).
\end{equation}
Thus, using the results of (\ref{meanq1}), the averages for $x_{1}$ are found to be
\begin{equation}
    \langle x_1 \rangle \:=\: 0, \qquad
    \langle x_1^2\rangle \:=\:  \frac{1}{4}\left[A_{\mathrm{c}}^2 \:+\: \left(\frac{m_2}{M}\right)^2 A_{\mathrm{r}}^2\right],
\end{equation}
with variance given by the following expression
\begin{equation}
    \Delta x_1 \:=\: \frac{1}{2} \sqrt{A_{\mathrm{c}}^2 \:+\: \left(\frac{m_2}{M}\right)^2 A_{\mathrm{r}}^2}.
\end{equation}
The symmetric properties of the system allow us to easily deduce $\Delta x_2$ by simply swapping the indices $1\rightarrow 2$ in the variables appearing in the expression for $\Delta x_1$.

Repeating the previous steps for the momenta allows to express $p_{1,2}$ in terms of the relative and centre of mass coordinate momenta defined as $p_{\mathrm{r}} \:=\: \mu \dot{x}_{\mathrm{r}}$ and $p_{\mathrm{c}} \:=\: M \dot{x}_{\mathrm{c}}$ respectively.
Hence, the time averages and variances of the momenta are 
\begin{eqnarray}
    \langle p_1 \rangle \:&=\: 0,\\
    \langle p_1^2 \rangle \:&=\: \frac{1}{4} \left[\mu^2 \omega_{\mathrm{r}}^2 A_{\mathrm{r}}^2 \:+\: \left(\frac{m_1}{M}\right)^2 M^2 \omega_{\mathrm{c}}^2 A_{\mathrm{c}}^2\right],\\
    \Delta p_1 \:&=\: \frac{1}{2} \sqrt{\mu^2 \omega_{\mathrm{r}}^2 A_{\mathrm{r}}^2 \:+\: \left(\frac{m_1}{M}\right)^2 M^2 \omega_{\mathrm{c}}^2 A_{\mathrm{c}}^2},
\end{eqnarray}
for the first oscillator. 
Once more, the results for the second block are obtained by swapping the subscripts $1 \rightarrow 2$.

Using the scaled variables $\overline{x}_i$ and $\overline{p}_i$ defined in \eref{scaled_xp}, enables us to find the dimensionless uncertainty product, $ \Delta \overline{x}\Delta \overline{p}$, as
\begin{equation} \label{x111}
     \Delta \overline{x}_1 \Delta \overline{p}_1 \:=\: \frac{1}{4}\sqrt{1 \:+\: \left(\frac{m_2}{M}\right)^2 \frac{A_{\mathrm{r}}^2}{A_{\mathrm{c}}^2}}\sqrt{\left(\frac{m_1}{M}\right)^2 \:+\: \frac{\mu^2 \omega_{\mathrm{r}}^2 A_{\mathrm{r}}^2}{M^2 \omega_{\mathrm{c}}^2 A_{\mathrm{c}}^2}  },
\end{equation}
and 
\begin{equation}\label{x112}
     \Delta \overline{x}_2 \Delta \overline{p}_2 \:=\: \frac{1}{4} \sqrt{1 \:+\: \left(\frac{m_1}{M}\right)^2 \frac{A_{\mathrm{r}}^2}{A_{\mathrm{c}}^2}}\sqrt{\left(\frac{m_2}{M}\right)^2 \:+\: \frac{\mu^2 \omega_{\mathrm{r}}^2 A_{\mathrm{r}}^2}{M^2 \omega_{\mathrm{c}}^2 A_{\mathrm{c}}^2}},
\end{equation}
for the first and second oscillator respectively.

We note that these expressions are symmetric under particle exchange and comprise of three dimensionless ratios involving the masses, the amplitudes and the frequencies of the oscillators. The extra term involving the masses in \eref{x111} is what sets it apart from the uncertainty product of the system of coupled oscillators with identical masses described by \eref{x12}, where only the other two dimensionless ratios appeared. This may be surprising, since both systems have the same number of fundamental dimensionful variables. The difference in the two expressions can be ascribed to the choice of coordinates of \eref{xrc} employed to decouple the equations of motion: since the expression for the centre of mass coordinate is weighted by the masses, this ``artificially'' introduces the extra term, $(m_i/M)^2$ with $i \:=\: 1,2$  in the uncertainty for momentum $\Delta \overline{p}_i$. Such a detail, however, is beyond the capabilities of dimensional analysis to predict.

\subsection{Quantum oscillators}
We now turn our attention to the quantum description of a system comprising of two distinguishable coupled harmonic oscillators with masses $m_1$ and $m_2$. The Hamiltonian of the system reads
\begin{equation}
\fl   \hat{H} \:=\: \frac{\hat p_1^2}{2m_1} \:+\: \frac{\hat p_2^2}{2m_2} \:+\: \frac{1}{2}m_1 \omega^2 x_1^2 \:+\: \frac{1}{2}m_2 \omega^2 x_2^2 \:+\: \frac{1}{2}k(x_1 \:-\: x_2)^2.
\end{equation}
As with the classical counterpart, the Hamiltonian here becomes separable under the Jacobi coordinate transformation of \eref{xrc}, i.e. 
\begin{eqnarray} \label{hami}
    \hat{H} \:&=\: \:-\: \frac{\hbar^2}{2\mu}\frac{\partial^{2} }{\partial x_{\mathrm{r}}^{2}} \:+\: \frac{1}{2}\mu \omega_{\mathrm{r}} x_{\mathrm{r}}^2 
    \:-\: \frac{\hbar^2}{2M}\frac{\partial^{2} }{\partial x_{\mathrm{c}}^{2}} \:+\: \frac{1}{2}M \omega_{\mathrm{c}} x_{\mathrm{c}}^2 \\
    \:&=\: \hat{H}_{\mathrm{r}} \:+\: \hat{H}_{\mathrm{c}},
\end{eqnarray}
where the meaning of the variables is the same as in Section \ref{classical}. The Hamiltonian \eref{hami} describes two single particle harmonic oscillators whose solutions are given by~\eref{ho}.

As before, the total wavefunction of the system is the product of the single-particle wavefunctions and the expectation values for $x_{\mathrm{r}}$ and $x_{\mathrm{c}}$ and the respective momentum operators are simply the one-particle results of \eref{align1} and \eref{align2}. With this information, we may evaluate the expectation values of the original spatial variables of the problem and the respective momenta:
\begin{eqnarray}
\fl    \langle x_1 \rangle \:&=\: 0,\qquad
    \langle x_1^2 \rangle \:=\: \frac{1}{4}\left[\frac{(2n_{\mathrm{c}} \:+\: 1)\hbar}{M\omega_{\mathrm{c}}} \:+\: \left(\frac{m_2}{M}\right)^2 \frac{(2n_{\mathrm{r}} \:+\: 1)\hbar}{\mu \omega_{\mathrm{r}}}\right],\\
\fl    \langle p_1\rangle \:&=\: 0,\qquad
    \langle p_1^2 \rangle \:=\: \frac{1}{4}\left[(2n_{\mathrm{r}} \:+\: 1)\mu \hbar \omega_{\mathrm{r}} \:+\: \left(\frac{m_1}{M}\right)^2(2n_{\mathrm{c}} \:+\: 1)M\hbar \omega_{\mathrm{c}}\right],
\end{eqnarray}
and likewise for the second oscillator variables $x_2$ and $p_2$.
\par In accordance with definition \eref{scaled_xp}, we introduce the variable $A_{n}$ to represent the classical turning point associated with the energy $E_n$ of the quantum mechanical harmonic oscillator. $A_n$ is as defined in \eref{amp}, where now $m$ is either $\mu$ or $M$ for $A_{n_{\mathrm{r}}}$ or $A_{n_{\mathrm{c}}}$ respectively.
This allows to find an expression for the dimensionless uncertainty product of position and momentum as
\begin{equation} \label{different}
    \Delta \overline{x}_1 \Delta \overline{p}_1 \:=\: \frac{1}{4} \sqrt{1+\left(\frac{m_2}{M}\right)^2 \frac{A_{n_{\mathrm{r}}}^2}{A_{n_{\mathrm{c}}}^2}}\sqrt{\left(\frac{m_1}{M}\right)^2 \:+\: \frac{\mu^2 \omega_{\mathrm{r}}^2 A_{n_{\mathrm{r}}}^2}{M^2 \omega_{\mathrm{c}}^2 A_{n_{\mathrm{c}}}^2}},
\end{equation}
and 
\begin{equation}\label{different1}
     \Delta \overline{x}_2 \Delta \overline{p}_2 \:=\: \frac{1}{4} \sqrt{1 \:+\: \left(\frac{m_1}{M}\right)^2 \frac{A_{n_{\mathrm{r}}}^2}{A_{n_{\mathrm{c}}}^2}}\sqrt{\left(\frac{m_2}{M}\right)^2 \:+\: \frac{\mu^2 \omega_{\mathrm{r}}^2A_{n_{\mathrm{r}}}^2}{M^2 \omega_{\mathrm{c}}^2 A_{n_{\mathrm{c}}}^2}},
\end{equation}
for the first and second oscillator respectively, where both canonical variables have been scaled by $A_{n_{\mathrm{c}}}$. To highlight the dependence on the principal quantum numbers, \eref{different} and \eref{different1} can be written explicitly as
\begin{eqnarray}
\fl    \Delta \overline{x}_1 \Delta \overline{p}_1 \:=\: \frac{1}{4} \sqrt{1 \:+\: \left(\frac{m_2}{M}\right)^2 \frac{(2n_{\mathrm{r}} \:+\: 1) M \omega_\mathrm{c}}{(2n_{\mathrm{c}} \:+\: 1) \mu \omega_\mathrm{r}}}  \sqrt{\frac{(2n_{\mathrm{r}} \:+\: 1) \mu \omega_{\mathrm{r}}}{(2n_{\mathrm{c}} \:+\: 1) M \omega_{\mathrm{c}}} \:+\: \left(\frac{m_1}{M}\right)^2}.
\end{eqnarray}
\begin{figure}
\centering
\subfloat[\label{fig:sfig12}]{%
  \includegraphics[width=0.3\columnwidth]{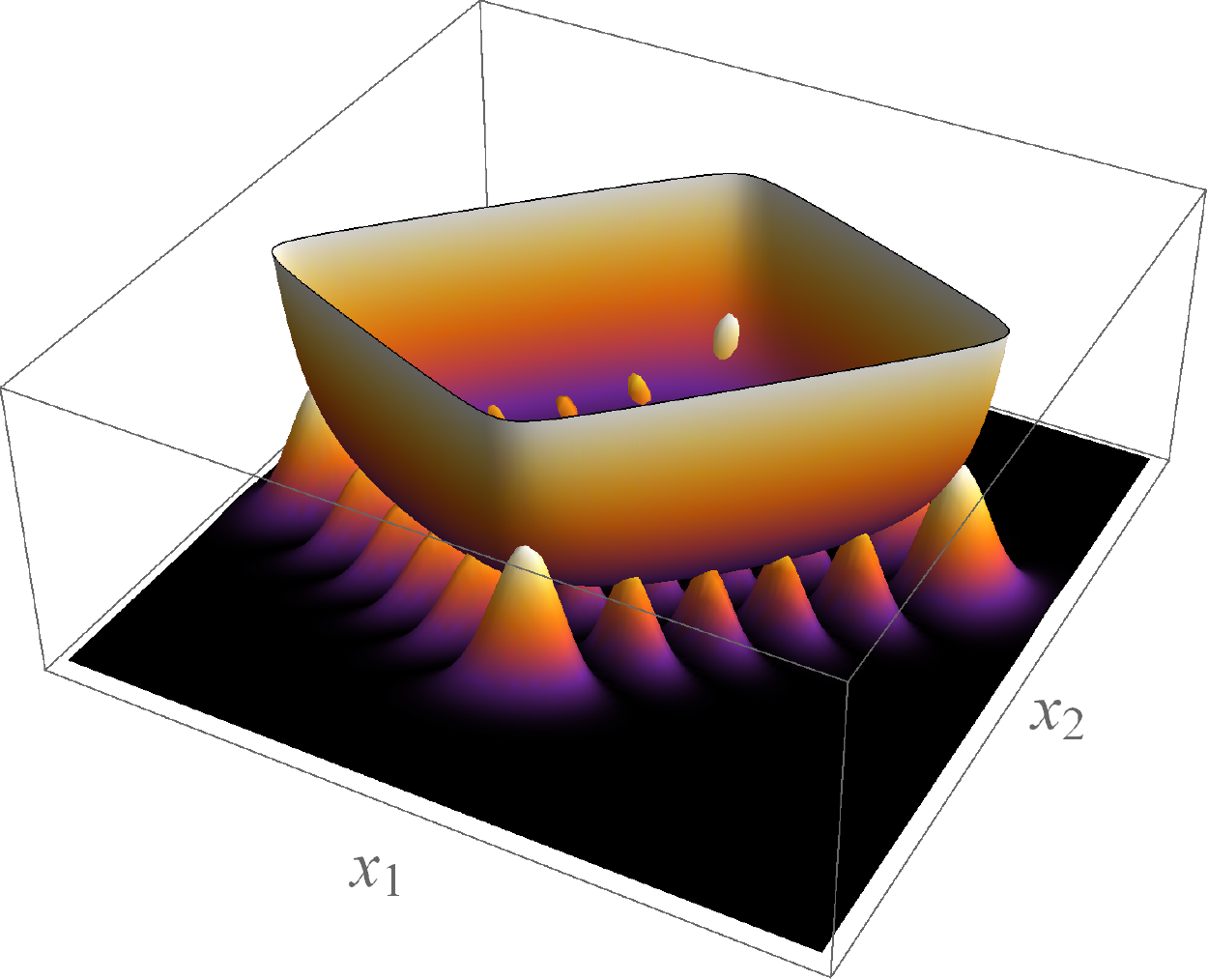}%
}
\subfloat[\label{fig:sfig28}]{%
  \includegraphics[width=0.3\columnwidth]{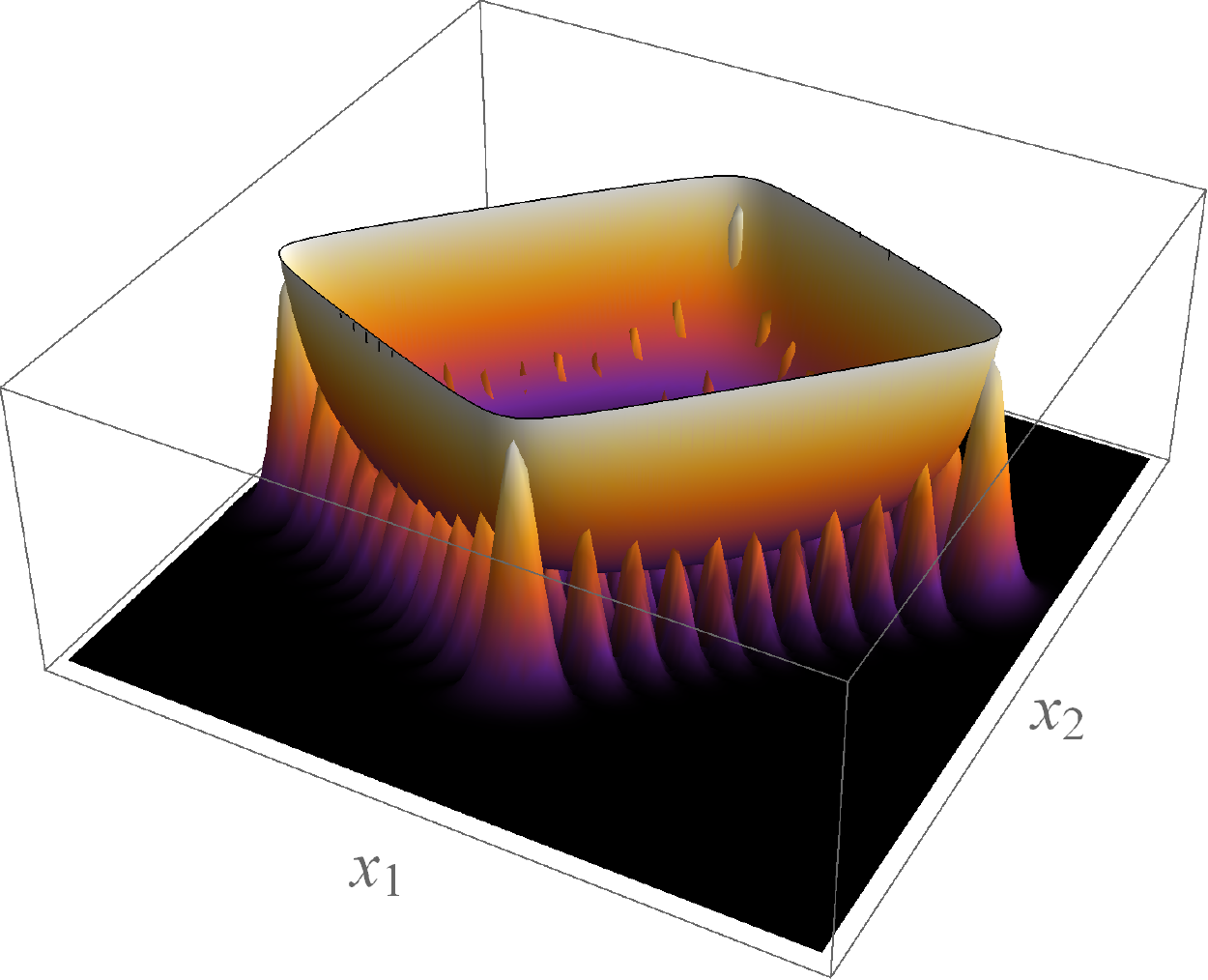}%
}
\subfloat[\label{fig:sfig29}]{%
  \includegraphics[width=0.3\columnwidth]{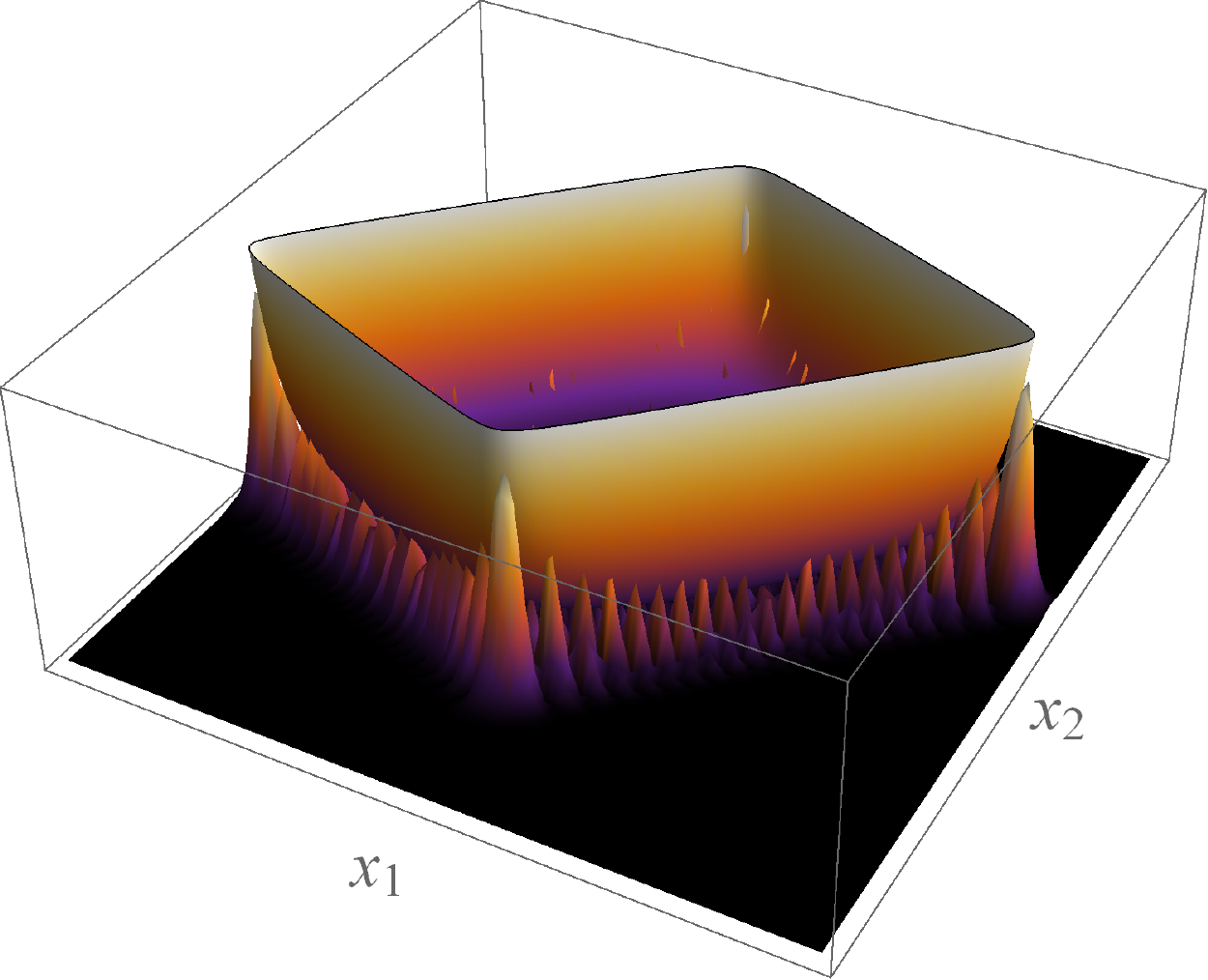}
}
\caption{Classical $\rho_{\mathrm{CL}}(x_1, x_2)$ and quantum $\rho_{\mathrm{QM}} = |\psi_n(x_1, x_2)|^2$ probability densities for coupled harmonic oscillators with different masses vs the spatial variables $x_{1}$ and $x_2$ for values of $n$ given by (a) $n \:=\: 5$, (b) $n \:=\: 10$, (c) $n \:=\: 20$.}\label{fig_6}

\end{figure}
Expectedly, the quantum uncertainty relations \eref{different} and \eref{different1} match the classical ones given by \eref{x111} and \eref{x112}. Furthermore, the considerations made regarding the overall effect of coupling on the uncertainty product (see end of Section \ref{CO_m}) and the difference to the results obtained in Section \ref{equal_masses} also apply to the expressions derived above. \par The graphical representations of the 1D classical and quantum probability densities are analogous to the ones presented in Figures \ref{fig_two} and \ref{fig_three}. Figure \ref{fig_6} offers a comparison between the quantum probability density, $\rho_{\mathrm{QM}} \:=\: |\Psi_n(x_1, x_2)|^2$ and the classical counterpart $ \rho_{\mathrm{CL}}(x_1, x_2)$ which are plotted as functions of the original variables of the problem for various values of $n$. Notably, the shape of the classical distribution is that of a skewed rectangular well which is not diagonal as the one in Figure \ref{fig_one}. It is evident however that the quantum and classical distributions converge in a locally averaged sense for high values of the principal quantum number $n$ in both cases.

\section{Two particles in a box coupled via contact potential}
The final set-up we consider is a system of two particles in an infinite potential well which are allowed to interact via a contact potential of the form
\begin{equation} \label{contact}
    V(x)=\left\{\begin{array}{ll}{\lambda \delta\left(x_1 \:-\: x_2\right)} & 0 < x_1,x_2 <L \\ {\infty} & {\mbox { otherwise }}\end{array}\right. ,
\end{equation}
where $L$ is the dimension of the box and the coefficient~$\lambda$, which has dimensions of $\rm M L^2 T^{-2}$, determines the strength of the interaction. For similar treatments of this set-up without uncertainty considerations, the reader is referred to \cite{trap, martin, pedram, L1, L2} or to \cite{belloni} for a treatment with uncertainty relations.

\subsection{Classical case}
Starting with the classical case, the Lagrangian of the system can be written as
\begin{equation} \label{lag}
    \mathcal{L} \:=\: \frac{1}{2} m_1\dot{x}_1^2 \:+\: \frac{1}{2}m_2 \dot{x}_2^2 - V,
\end{equation}
where $V$ is defined in \eref{contact} and, in general, $m_1 \neq m_2$.
Once again the equations of motion are decoupled via the Jacobi coordinates:
\begin{equation} \label{J_1}
     x_\mathrm{r} \:=\: x_1 \:-\: x_2, \quad \quad \quad  x_\mathrm{c} \:=\: \frac{m_1 x_1 \:+\: m_2 x_2}{M},
\end{equation}
where $M \:=\: m_1 \:+\: m_2$ is the total mass of the system of particles. Proceeding with this variable transformation allows (\ref{lag}) to be written as
\begin{equation}
      \mathcal{L} \:=\: \frac{1}{2} M\dot{x}_{\mathrm{c}}^2+ \frac{1}{2}\mu \dot{x}_{\mathrm{r}}^2 \:-\: \lambda \delta\left(x_{\mathrm{r}}\right),
\end{equation}
where $\mu \:=\: (m_1 m_2) /M$ is the reduced mass and the expression for $\mathcal{L}$ is valid within the spatial domain $0<x_1,x_2<L$. The equation of motion for the center of mass coordinate is trivial $M \ddot{x}_{\mathrm{c}} \:=\: 0$, as expected for a classical particle confined in a one-dimensional potential box.

Let us now evaluate the classical averages relevant to this set-up. Since $0<x_1,x_2<L$ then the spatial domain for the centre of mass coordinate is $0<x_{\mathrm{c}}<L$. 
Hence, the time averages $\langle x_{\mathrm{c}} \rangle$ and $\langle p_{\mathrm{c}} \rangle $ for position and momentum are simply
\begin{eqnarray}
    \langle x_{\mathrm{c}} \rangle \:&=\: \frac{L}{2},\quad \quad \quad \langle x_{\mathrm{c}}^2 \rangle \:=\: \frac{L^2}{3},\\
     \langle p_{\mathrm{c}} \rangle \:&=\: 0, \quad \quad \quad \langle p_{\mathrm{c}}^2 \rangle \:=\: 2ME_{\mathrm{c}} ,
\end{eqnarray}
where $E_{\mathrm{c}}$ is the energy of the particle.
For the relative coordinate, the equation of motion involves the derivative of the delta-function which is remarkably challenging to compute analytically. As a simplification, however, since we are in the classical regime where no tunnelling can occur, we can assume that the potential acts as a solid wall at $x_{\mathrm{r}} =0$. Essentially, in each half of the box we are dealing with a classical free particle confined to move in a limited region of space with the condition $-L<x_{\mathrm{r}} <L$. Hence, the averages will again be those of the one-particle case for a box of length $2L$ with the condition that $x_{\mathrm{r}} \neq 0$ \cite{R6, usha}. 

We find the averages to be 
\begin{eqnarray}
    \langle x_{\mathrm{r}} \rangle \:&=\: 0,\quad \quad \quad \langle x_{\mathrm{r}}^2 \rangle \:=\: \frac{L^2}{3},\\
     \langle p_{\mathrm{r}} \rangle \:&=\: 0, \quad \quad \quad \langle p_{\mathrm{r}}^2 \rangle \:=\: 2\mu E_\mathrm{}.
\end{eqnarray}
Switching back to the original variables, the classical averages of the spatial coordinate $x_1$ and the respective momentum $p_1$ are found to be
\begin{eqnarray}
   \langle x_1 \rangle \:&=\: \frac{L}{2}, \qquad
   \langle x_1^2 \rangle \:&=\: \frac{L^2}{3}\left[1 \:+\: \left(\frac{m_2}{M}\right)^2\right], \label{113}\\
   \langle p_1 \rangle \:&=\: 0,\qquad
   \langle p_1^2 \rangle \:&=\: 2 \mu E_{\mathrm{r}} \:+\: \left(\frac{m_1}{M}\right)^2 2 M E_{\mathrm{c}},\label{116}
\end{eqnarray}
and similarly for the second particle:
\begin{eqnarray}
   \langle x_2 \rangle \:&=\: \frac{L}{2}, \qquad
   \langle x_2^2 \rangle \:&=\: \frac{L^2}{3}\left[1 \:+\: \left(\frac{m_1}{M}\right)^2 \right],\label{118} \\
   \langle p_2 \rangle \:&=\: 0,\qquad
   \langle p_2^2 \rangle \:&=\: 2 \mu E_{\mathrm{r}} \:+\: \left(\frac{m_2}{M}\right)^2 2 M E_{\mathrm{c}}.\label{120}
\end{eqnarray}
To find the dimensionless uncertainty product $\Delta x\Delta p$, we use a similar definition to that of (\ref{scaled_xp}), introducing the following scaled quantities
\begin{equation} \label{def}
    \Delta \overline{x} \:=\: \frac{\Delta x}{L}, \quad \quad \quad \Delta \overline{p} \:=\: \frac{\Delta p}{\sqrt{2mE}}.
\end{equation}

Hence, after finding the variances using the results \eref{113} to \eref{120} and expressing them in terms of the canonical variables from \eref{def}, one finds the dimensionless uncertainty product of position and momentum to be
\begin{equation} \label{box1}
     \Delta \overline{x}_1  \Delta \overline{p}_1 \:=\: \sqrt{\frac{1}{12}} \sqrt{1 \:+\: \left(\frac{2 m_2}{M}\right)^2}\sqrt{1 \:+\: \left(\frac{m_1}{M}\right)^2\frac{M E_{\mathrm{c}}}{\mu E_{\mathrm{r}}}}.
\end{equation}
Since the system is symmetric under exchange of particles, the product $\Delta x_2 \Delta p_2$ can be easily deduced from \eref{box1} by exchanging the indices $1$ and $2$:
\begin{equation}\label{box2}
    \Delta \overline{x}_2  \Delta \overline{p}_2 \:=\: \sqrt{\frac{1}{12}} \sqrt{1 \:+\:  \left(\frac{2 m_1}{M}\right)^2}\sqrt{1 \:+\: \left(\frac{m_2}{M}\right)^2\frac{M E_{\mathrm{c}}}{\mu E_{\mathrm{r}}}}.
\end{equation}
We can see that \eref{box1} (and equivalently \eref{box2}) only involves dimensionless ratios of the masses and energies of the two classical particles. Comparing these results to the uncoupled case relation, which is given by \eref{3case}, we see how the two expressions do not simply coincide when $m_1 \:=\: m_2$ and $E_{\mathrm{c}} \:=\: E_{\mathrm{r}}$. In fact, in the latter case we obtain exactly double the uncertainty product of the one-particle case. This can be ascribed to the fact that this is still a coupled system of particles.\par Hence, as we have also concluded in the other examples considered so far, introducing a coupling potential increases the uncertainty product  $ \Delta \overline{x}  \Delta \overline{p}$ from its minimum value given by \eref{3case} through an additional dimensionless ratio of parameters.

\subsection{Quantum case}
The quantum mechanical counterpart of the system under consideration is described by the following Hamiltonian operator
\begin{equation} \label{H_box}
     \hat{H} \:=\: \frac{\hat{p}_1^2}{2m_1} \:+\: \frac{\hat{p}_2^2}{2m_2} \:+\: \lambda \delta \left(x_1 \:-\: x_2\right) .
\end{equation}
It is important to point out that a quantum mechanical system  coupled by a contact interaction of the kind described above can only consist of bosonic states. This follows from the Pauli exclusion principle: since fermions cannot be at the same point in space at the same instant in time, a contact potential would not affect fermionic states \cite{marsiglia}.

Proceeding with the analysis, \eref{H_box} can be decoupled by transforming to the same Jacobi coordinates defined in \eref{J_1} for the classical case. The resulting Hamiltonian is separable in terms of the centre of mass and relative coordinate:
\begin{equation}
     \hat{H} = -\frac{\hbar^2}{2M}\frac{\partial^{2} }{\partial x_{\mathrm{c}}^{2}} -\frac{\hbar^2}{2\mu}\frac{\partial^{2} }{\partial x_{\mathrm{r}}^{2}} +\lambda \delta \left(x_{\mathrm{r}}\right).
\end{equation}
The solution to the centre of mass part of the Hamiltonian is that of a single particle harmonic oscillator with mass $M$ in a one-dimensional box of length $L$, i.e. $0<x_{\mathrm{c}}<L$. The expectation values for position and momentum are then given by
\begin{eqnarray}
 \langle x_{\mathrm{c}} \rangle &= \frac{L}{2}, \qquad
    &\langle x_{\mathrm{c}}^2 \rangle = L^2 \left(\frac{1}{3}-\frac{1}{2n_{\mathrm{c}}^2\pi^2}\right),\\
  \langle p_{\mathrm{c}}\rangle &= 0, \qquad
    &\langle p_{\mathrm{c}}^2 \rangle = \left(\frac{\pi \hbar n_{\mathrm{c}}}{L}\right)^2 = 2 M E_{n_{\mathrm{c}}}.
\end{eqnarray}
\begin{figure}
    \centering
    \includegraphics[scale=0.5]{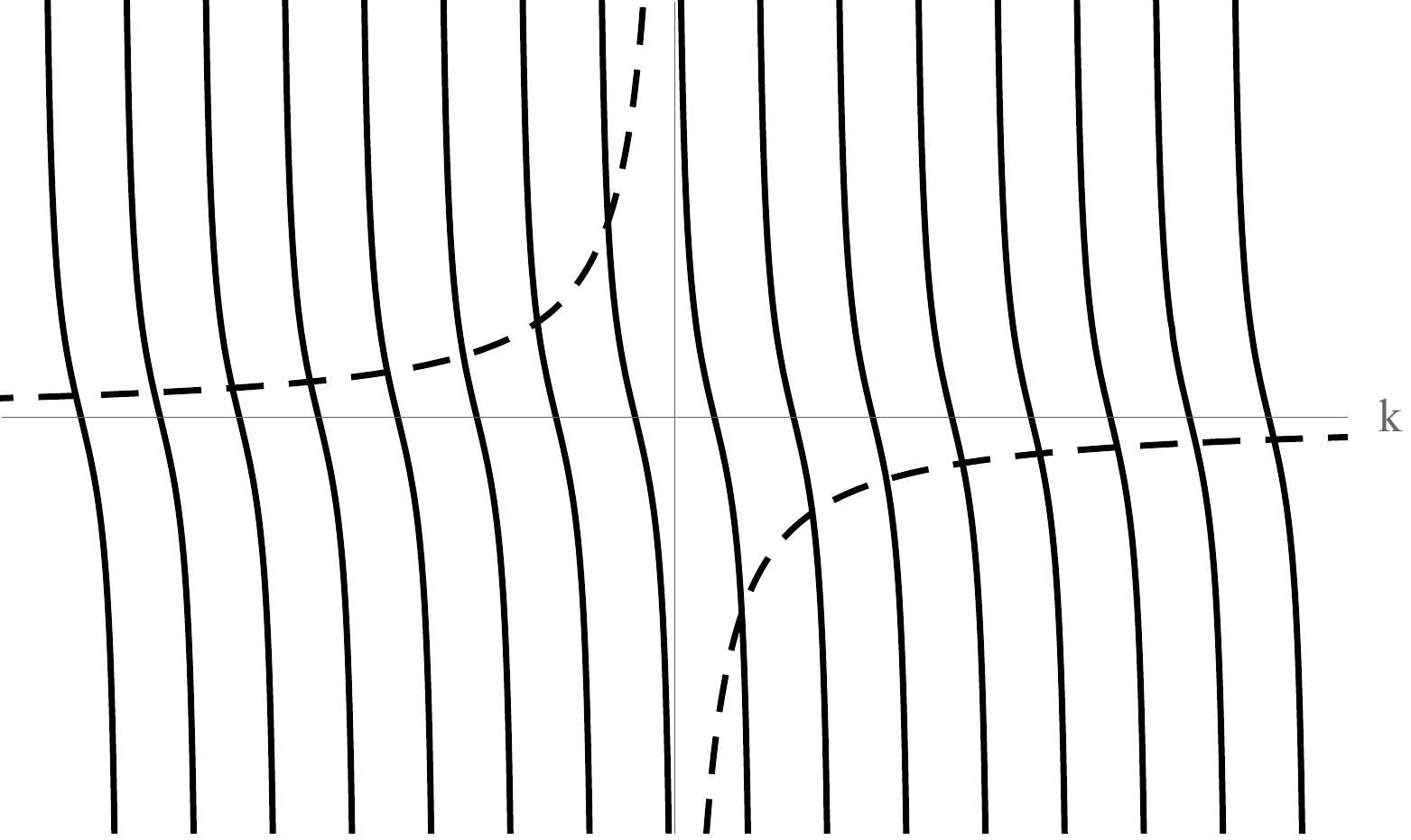}
    \caption{Plot of $\cot{\left(k_{n_{\mathrm{r}}}\right)}$ (solid lane) vs $-1/k_{n_{\mathrm{r}}}$ (dashed line) for a box of length $L=5$ and for $\mu \lambda /\hbar^2 =1$.}
    \label{fig_9}
\end{figure}
\begin{figure}
    \centering
    \includegraphics[scale=0.5]{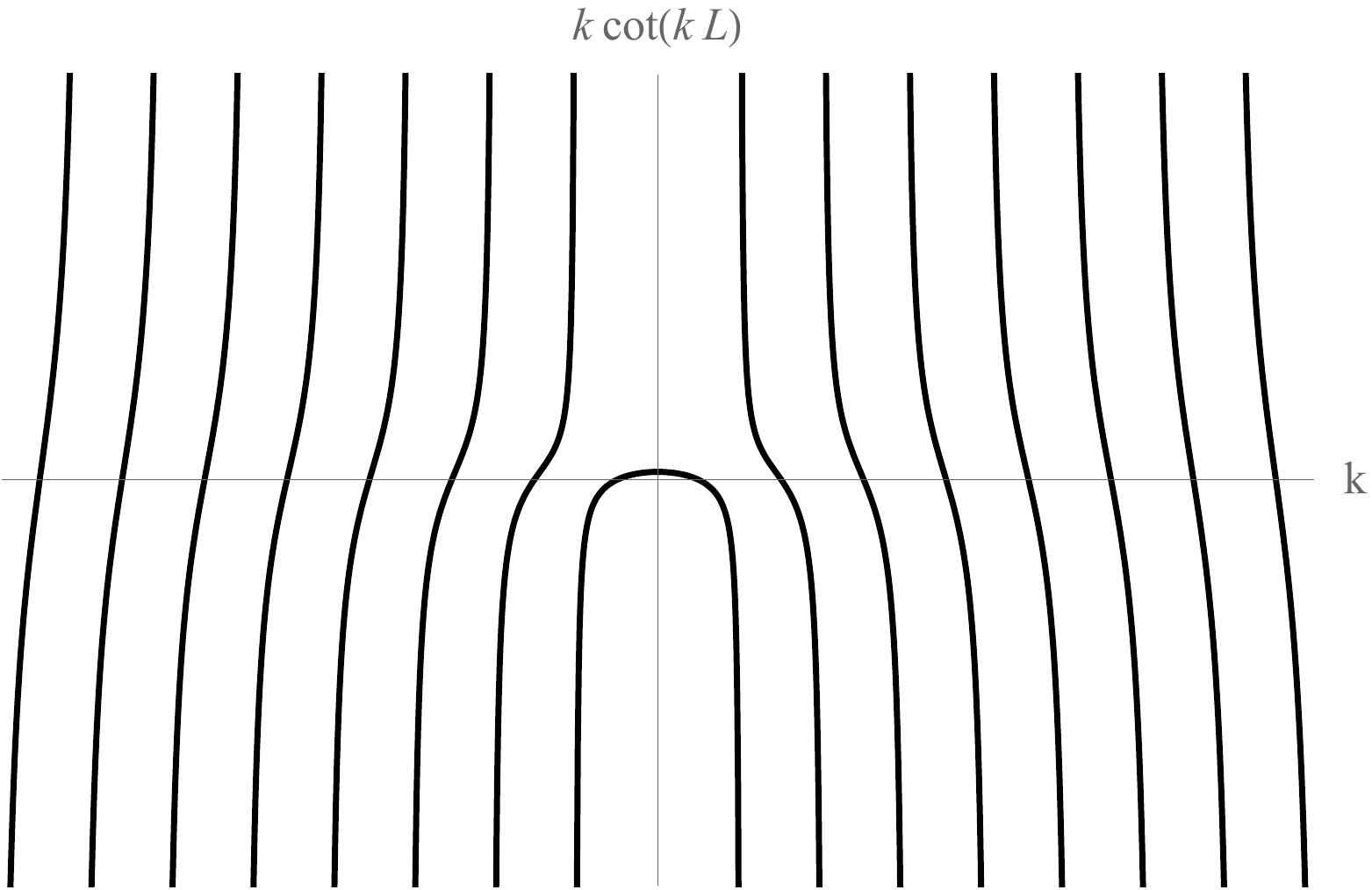}
    \caption{Plot of the self-consistent equation (\ref{k_r}) vs $k_{n_{\mathrm{r}}}$ for a box of length $L=5$ within a range $-10<k_{n_{\mathrm{r}}}<10$.}
    \label{fig_8}
\end{figure}
In addition to the box boundary conditions, the presence of delta-function potential in the relative coordinate Hamiltonian imposes two sets of boundary conditions on the wavefunction at the location of the delta-function potential. These are the continuity of the wavefunction
\begin{equation} 
        \left.\psi\right|_{x_{1}=x_{2}+0}=\left.\psi\right|_{x_{1}=x_{2}-0},\label{continuity}
\end{equation}
and the discontinuity of its first derivative
\begin{eqnarray}
\left.\frac{d \psi_{n_{\mathrm{r}}}(x_{\mathrm{r}})}{d x_{\mathrm{r}}}\right|_{0+\epsilon}-\left.\frac{d \psi_{n_{\mathrm{r}}}(x_{\mathrm{r}})}{d x_{\mathrm{c}}}\right|_{0-\epsilon} &=\frac{2 \mu}{\hbar^{2}} \int_{0-\epsilon}^{0+\epsilon} V(x_{\mathrm{r}}) \psi_{{n_{\mathrm{r}}}}(x_{\mathrm{r}}) d x_{\mathrm{r}}  \\
&=\frac{2 \mu \lambda}{\hbar^{2}} \psi_{n_{\mathrm{r}}}\left(0\right).\label{discontinuity}
\end{eqnarray}
The latter is obtained by integrating the Schr\"odinger equation with Hamiltonian operator specified by (\ref{H_box}) over the small
interval $\left(0-\epsilon, 0+\epsilon\right)$  \cite{bera, jog}.

From \cite{martin,pedram}, we can see that for a particle confined in a box of length $2L$, i.e. $-L<x_{\mathrm{r}}<L$, with a $\delta$-function potential at position $x_{\mathrm{r}} = pL$, where $-1<p<1$, the wavefunction can be written as
\begin{equation}
    \psi_{n_\mathrm{r}}(x_\mathrm{r})=\left\{\begin{array}{ll}{A \sin \left(k_{n_\mathrm{r}} (x_\mathrm{r}+L)\right)} & {(-L \leq x_\mathrm{r} \leq pL)} \\ {B \sin \left[k_{n_\mathrm{r}}(x_\mathrm{r}-L)\right]} & {(pL \leq x_\mathrm{r} \leq L)}\end{array}\right. ,
\end{equation}
where $k_{n_\mathrm{r}}=\sqrt{2\mu E_\mathrm{r}}/\hbar$. Moreover, the continuity condition  \eref{continuity} of the wave function at $x_{\mathrm{r}} =pL $ gives 
\begin{equation}
    \frac{A}{B}=\frac{\sin \left[k_{n_{\mathrm{r}}}L(p-1)\right]} { \sin \left[k_{n_{\mathrm{r}}}L(p+1)\right]}.
\end{equation}
If $p=0$, this condition yields $A=-B$. Applying now the discontinuity condition of~\eref{discontinuity} allows us to obtain a quantization relation for wavenumber $k_{n_{\mathrm{r}}}$, i.e. 
\begin{equation}
    k_{n_{\mathrm{r}}} \sin \left(2k_{n_{\mathrm{r}}} L\right)=\frac{2 \mu \lambda}{\hbar^{2}} \sin \left[k_{n_{\mathrm{r}}} L(p-1) \right] \sin \left[k_{n_{\mathrm{r}}} L (p+1)\right].
\end{equation}
Setting $p=0$ yields the following expression
\begin{equation} \label{k_r}
    k_{n_{\mathrm{r}}}\cot{\left(k_{n_{\mathrm{r}}}L\right)} = - \frac{\mu \lambda}{\hbar^2},
\end{equation}
which is a self-consistent relation for $k_{n_{\mathrm{r}}}$ and cannot be solved to give a simple analytical answer for the wavenumber. This is shown in Figure \ref{fig_9} where the intersection points between the curves $y=\cot{(k_{n_{\mathrm{r}}} L)}$ and $y=-1/k_{n_{\mathrm{r}}}$ represent the solutions to~\eref{k_r}. Clearly, these solutions are not spaced in a regular manner and a numerical approach becomes necessary.
Plotting~\eref{k_r} against $k_{n_{\mathrm{r}}}$ however gives us insight into the way the wavefunction changes over the length of the box. For a box of length $L=5$ and within a range $-10< k_{n_{\mathrm{r}}}<10$, the behaviour of the function is shown in Figure \ref{fig_8}. \par We note that for large values of the ordinate, the lines become parallel and equally spaced, suggesting that in the high quantum number limit, the $k_{n_{\mathrm{r}}}$ values satisfying the self-consistent equation will repeat periodically. After normalizing the wavefunctions, the expectation values for the relative spatial coordinate, $x_{\mathrm{r}}$, and momentum are evaluated to be
 \begin{eqnarray}
     \label{x_r2}
 \langle x_{\mathrm{r}} \rangle &= 0, \quad \quad  \langle x_{\mathrm{r}}^2 \rangle = \frac{2 k_{n_{\mathrm{r}}} L^3}{6 k_{n_{\mathrm{r}}} L-3 \sin (2 k_{n_{\mathrm{r}}} L)}-\frac{1}{2 k_{n_{\mathrm{r}}}^2},\\
    \langle p_{\mathrm{r}} \rangle &=0, \quad \quad 
    \langle p_{\mathrm{r}}^2 \rangle = \hbar^2 k_{n_{\mathrm{r}}}^2 = 2\mu E_{n_{\mathrm{r}}}.
\end{eqnarray}
The expression for $\langle x_{\mathrm{r}}^2 \rangle$ in \eref{x_r2} seems quite complicated at a first glance  but reduces to the familiar $L^2/3$ in the limit of large $k_{n_{\mathrm{r}}}$. This corresponds to the classical regime:
\begin{equation}
     \lim_{k_{n_{\mathrm{r}}}\to\infty} \langle x_\mathrm{r}^2 \rangle \rightarrow \frac{L^2}{3}.
\end{equation}
Hence, the expectation values of $x_1$ and $p_1$ are found to be
\begin{eqnarray}
    \langle x_1 \rangle &= \frac{L}{2}, \qquad
    \langle x_1^2 \rangle &= L^2 \left(\frac{1}{3}-\frac{1}{2n_c^2\pi^2}\right)+\left(\frac{m_2}{M}\right)^2 \langle x_{\mathrm{r}}^2 \rangle,\\
    \langle p_1 \rangle &= 0,\qquad 
    \langle p_1^2 \rangle &= 
     2\mu E_{\mathrm{r}} +\left(\frac{m_1}{M}\right)^2 2M E_{\mathrm{c}},
\end{eqnarray}
and similarly for the second particle.
Using the scaled variables of \eref{def} we can then evaluate the dimensionless uncertainties in position and momentum as
\begin{eqnarray}
    \Delta \overline{x}_1 &=\sqrt{\frac{1}{12}-\frac{1}{2 n_{\mathrm{c}}^2 \pi^2}+\left(\frac{m_2}{M}\right)^2 \frac{\langle x_{\mathrm{r}}^2 \rangle}{L^2}},\\
    \Delta \overline{p}_1 &= \sqrt{1+\left(\frac{m_1}{M}\right)^2\frac{M E_{\mathrm{c}}}{\mu E_{\mathrm{r}}}}. 
\end{eqnarray}
Taking the limit for large principal quantum number $n_{\mathrm{c}}$ and wavenumber $k_{n_{\mathrm{r}}}$, we step in the classical regime and obtain an equivalent result to \eref{box1}, i.e. 
\begin{eqnarray}
   \lim_{n_{\mathrm{c}},  k_{n_{\mathrm{r}}}\to\infty}  \Delta \overline{x}_1\Delta \overline{p}_1 &= \sqrt{\frac{1}{12}} \sqrt{1+\left(\frac{2 m_2}{M}\right)^2} 
   \sqrt{1+\left(\frac{m_1}{M}\right)^2\frac{M E_{\mathrm{c}}}{\mu E_{\mathrm{r}}}} .
\end{eqnarray}
This is in complete agreement with the classical result of  \eref{box1}.

As it is not possible to obtain a value for the wavenumber $k_{n_{\mathrm{r}}}$ in closed form, we will not offer a graphical comparison between quantum and classical probability densities here. The reader is referred to \cite{usha} for a comparison of probability density distributions for the simple one-particle infinite square well.

\section{Conclusions}

Dimensional analysis is a powerful tool for deriving robust results for a host of situations, and the correspondence between the quantum and classical domains is no exception. The Buckingham-$\pi$ theorem of dimensional analysis provides us with a fundamental insight regarding the so-called classical limit: its definition cannot involve taking the limit of a dimensionful quantity such as $\hbar$ if it is to be physically meaningful. Even in cases where $\hbar \to 0$ is a shorthand for some quantum number $n\to \infty$, it is not possible to recover classical mechanics by na\"ively setting $\hbar = 0$.

There is a stark divide between a classical world where $\hbar = 0$ and a quantum world viewed by an observer living in the classical limit. The classical probability density is not a limit of the wavefunction in the classical limit: indeed, the classical probability density can be recovered by ``smearing out" the quantum probability density (by an amount that decreases as quantum numbers increase). In a classical world, wavefunctions do not exist, but in a quantum world, classical scales are those in which the smearing required to match the two domains is small (i.e. wavenumbers are very large).  

In order to illustrate how the correspondence between classical and quantum realm fares in the context of dimensional analysis, we focused on quantum uncertainty, popularly regarded as a phenomenon with no classical analogue. For simple systems, the role of $\hbar$ in the classical limit is assumed by a characteristic action scale, as required by the Buckingham-$\pi$ theorem. For systems with more dimensionful parameters, however, uncertainty limits will in general acquire a more complicated dependence on dimensionless ratios. To demonstrate this, we examined two distinct systems: a coupled harmonic oscillator and a two-body particle-in-a-box set-up, and showed that the uncertainty bounds acquire additional dimensionless terms (and also converge in the limit of large quantum numbers). 
 
Dimensional analysis lends support to the idea that any comparison between quantum and  classical uncertainties is only possible if made in terms of dimensionless quantities, since the classical realm lacks a fundamental unit of action (much like Newtonian mechanics lacks a fundamental unit of speed).
Quantum numbers are not simply the discretized counterparts to the dimensionless ratios of the classical system; they arise precisely because of the existence of the fundamental unit $\hbar$. This highlights the fundamental conceptual distinction between quantum and classical mechanics: it is only through the tuning of \emph{dimensionless} parameters that a bridge between the two domains can be unambiguously built. This indicates that introducing uncertainty to a system is, in a sense, a one-way street. Uncertainty bounds can only exist alongside a dimensionful parameter that acts as a ``universal certainty limit" (just like relativity requires a universal speed limit). Once such a parameter is added to a theory, there is a well-defined limit in which the quantum world appears to be classical, but there is no limit that will return us to a truly classical world.

\end{document}